\documentclass[acmsmall,screen,nonacm]{acmart} 

\AtBeginDocument{%
  \providecommand\BibTeX{{%
    \normalfont B\kern-0.5em{\scshape i\kern-0.25em b}\kern-0.8em\TeX}}}

\usepackage{multirow}
\usepackage{longfbox}
\usepackage{graphicx}
\usepackage{amsmath}
\usepackage[american]{babel}
\usepackage{microtype}%
\usepackage{subcaption}
\usepackage[ruled,,linesnumbered ]{algorithm2e}
\usepackage{tcolorbox}
\usepackage{colortbl}
\usepackage{balance} 
\usepackage{enumitem} 
\usepackage{color}
\usepackage{url}

\usepackage{mwe}
\usepackage{wrapfig}

\newcommand{\method}{\textsc{CodeMI \xspace}}%
\newcommand{\methodnospace}{\textsc{CodeMI}}%

\usepackage{tabularx}

\newcount\DraftStatus  
\definecolor{darkgreen}{rgb}{0,0.5,0} 
\definecolor{purple}{rgb}{1,0,1} 
\definecolor{todocolor}{rgb}{0.9,0.1,0.1} 
\definecolor{fixcolor}{rgb}{0.1,0.7,0.3} 
\definecolor{wycolor}{rgb}{0.9,0.1,0.1} 
\definecolor{hycolor}{rgb}{0.7,0.7,0.3} 
\definecolor{zqcolor}{rgb}{0.79, 0.63, 0.86} 
\definecolor{ashgrey}{rgb}{0.7, 0.75, 0.71}
\definecolor{grey}{rgb}{0.6,0.6,0.6}
\definecolor{lightblue}{rgb}{0.76, 0.82, 0.94}

\newcommand{\nbc}[3]{\ifnum\DraftStatus=1
	{\colorbox{#3}{\bfseries\sffamily\scriptsize\textcolor{white}{#1}}}
	{\textcolor{#3}{\sf\small$\blacktriangleright$\emph{#2}$\blacktriangleleft$}}
\fi}

\newcommand{\draftnote}[2]{\ifnum\DraftStatus=1
	\marginpar{
		\tiny\raggedright
		\hbadness=10000
		\def\baselinestretch{0.8}
		\textcolor{#1}{\textsf{\hspace{0pt}#2}}}
\fi}

\makeatletter
\def\thickhline{%
  \noalign{\ifnum0=`}\fi\hrule \@height \thickarrayrulewidth \futurelet
   \reserved@a\@xthickhline}
\def\@xthickhline{\ifx\reserved@a\thickhline
               \vskip\doublerulesep
               \vskip-\thickarrayrulewidth
             \fi
      \ifnum0=`{\fi}}
\makeatother

\makeatletter
\def\old@comma{,}
\catcode`\,=13
\def,{%
  \ifmmode%
    \old@comma\discretionary{}{}{}%
  \else%
    \old@comma%
  \fi%
}
\makeatother

\newlength{\thickarrayrulewidth}
\renewcommand\footnotetextcopyrightpermission[1]{} 

\begin{document}

\title[]{Does Your Neural Code Completion Model Use My Code? A Membership Inference Approach}

%
\author{Yao Wan}
\email{wanyao@hust.edu.cn}
\affiliation{%
  \department{National Engineering Research Center for Big Data Technology and System, Services Computing Technology and System Lab, Cluster and Grid Computing Lab, School of Computer Science and Technology}
	\institution{Huazhong University of Science and Technology}
	\city{Wuhan}
	\country{China}
}
\author{Guanghua Wan}
\email{wanguanghua@hust.edu.cn}
\affiliation{%
  \department{National Engineering Research Center for Big Data Technology and System, Services Computing Technology and System Lab, Cluster and Grid Computing Lab, School of Computer Science and Technology}
	\institution{Huazhong University of Science and Technology}
	\city{Wuhan}
	\country{China}
}
\author{Shijie Zhang}
\email{shijie_zhang@hust.edu.cn}
\affiliation{%
  \department{National Engineering Research Center for Big Data Technology and System, Services Computing Technology and System Lab, Cluster and Grid Computing Lab, School of Computer Science and Technology}
	\institution{Huazhong University of Science and Technology}
	\city{Wuhan}
	\country{China}
}

\author{Hongyu Zhang}
\affiliation{%
	\institution{Chongqing University}
	\country{China}}
\email{hyzhang@cqu.edu.cn}
\author{Pan Zhou}
\affiliation{%
	\institution{Huazhong University of Science and Technology}
	\country{China}}
\email{panzhou@hust.edu.cn}
\author{Hai Jin}
\email{hjin@hust.edu.cn}
\affiliation{%
  \department{National Engineering Research Center for Big Data Technology and System, Services Computing Technology and System Lab, Cluster and Grid Computing Lab, School of Computer Science and Technology}
	\institution{Huazhong University of Science and Technology}
	\city{Wuhan}
	\country{China}
}

\author{Lichao Sun}
\affiliation{%
	\institution{University of Leigh}
	\country{USA}
}
\email{lis221@lehigh.edu}


\renewcommand{\shortauthors}{Wan et al.}

\begin{abstract}
Recent years have witnessed significant progress in developing deep learning-based models for automated code completion. Examples of such models include CodeGPT and StarCoder. These models are typically trained from a large amount of source code collected from open-source communities such as GitHub. Although using source code in GitHub has been a common practice for training deep-learning-based models for code completion, it may induce some legal and ethical issues such as copyright infringement. 
In this paper, we investigate the legal and ethical issues of current neural code completion models by answering the following question: \textit{Is my code used to train your neural code completion model?} 

To this end, we tailor a membership inference approach (termed \methodnospace) that was originally crafted for classification tasks to a more challenging task of code completion. 
In particular, since the target code completion models perform as opaque black boxes, preventing access to their training data and parameters, we opt to train multiple shadow models to mimic their behavior.
The acquired posteriors from these shadow models are subsequently
employed to train a membership classifier. 
Subsequently, the membership classifier can be effectively employed to deduce the membership status of a given code sample based on the output of a target code completion model.
We comprehensively evaluate the effectiveness of this adapted approach across a diverse array of neural code completion models,  (i.e.,  LSTM-based, CodeGPT, CodeGen, and StarCoder).
Experimental results reveal that the LSTM-based and CodeGPT models suffer the membership leakage issue, which can be easily detected by our proposed membership inference approach with an accuracy of 0.842, and 0.730, respectively.
Interestingly, our experiments also show that the data membership of current large language models of code, e.g., CodeGen and StarCoder, is difficult to detect, leaving amper space for further improvement.
Finally, we also try to explain the findings from the perspective of model memorization.

\end{abstract}





%


%

\maketitle

\section{Introduction}\label{sec_introduction}
Code completion, with its goal of providing automatic suggestions based on the input prompts (e.g., contextual partial code and natural-language descriptions), has the potential to notably enhance the programming productivity of developers.
Recently, many deep-learning-based approaches and tools have been developed for code completion.
Examples of these models include TabNine~\cite{tabnine}, CodeGPT~\cite{lu2021codexglue}, IntelliCode~\cite{intellicode}, as well as a range of recent \textit{Large Language Models} (LLMs), such as GitHub Copilot~\cite{copilot}, CodeGen~\cite{nijkamp2023codegen}, Code Llama~\cite{roziere2023code}, and StarCoder~\cite{li2023starcoder}.
While deep-learning-based approaches have achieved significant progress in code completion, the use of source code for model training may induce some legal and ethical concerns. 
This arises from the fact that numerous open-source projects on GitHub are governed by specific licenses that permit users to legally reuse, modify, and distribute the source code. To illustrate, the GNU GPLv3 license grants users the freedom to undertake nearly any action with the open-source project, except for distributing closed-source variants.
These legal and ethical issues have been amplified by the recent pre-trained code models (e.g, Codex~\cite{codex}, CodeGPT~\cite{lu2021codexglue}, and ChatGPT~\cite{chatgpt}) for code completion, which require a very large-scale code corpus for model pre-training.

In this paper, we aim to investigate the legal and ethical issues of current neural code completion models by answering the following question: 
\textit{Is my code used to train your neural code completion model?}
Exploring this question holds the promise of yielding benefits in two key application scenarios: 
safeguarding intellectual property and stealing private training data used by code completion models.

\noindentparagraph{\textbf{\textup{Scenario 1: Safeguarding intellectual property.}}}
In a typical real-world scenario, a service provider constructs a neural code completion model using an undisclosed dataset, which is consistently sourced from GitHub. This model is subsequently offered as a service, often referred to as ``\textit{machine learning as a service}'', accessible through a public API. 
Subsequently, the client user has the capability to employ their current partial code snippet as an input query to access the API, and in return, they receive a probability distribution for the next code token. 
Now, the user may have a vested interest in determining whether their code, which has been released online, was employed in the training dataset for the neural code completion models.

\noindentparagraph{\textbf{\textup{Scenario 2: Stealing private training data used by code completion models.}}}

Typically, current code completion models undergo training using vast GitHub code repositories, potentially containing an array of sensitive personal information~\cite{meli2019bad,niu2023codexleaks}. This training data may span from personally identifiable information (e.g., emails and social media accounts) to confidential data (e.g., SSNs and medical records) and even secret information (e.g., passwords, access keys, and PINs). Moreover, the training data may also contain data from both publicly available sources and conceivably private user-generated code~\cite{github-private}.
Recent studies~\cite{yang2024unveiling} have revealed that existing code language models possess the capability to memorize certain aspects of their training data, thereby potentially generating sensitive information.
In a malicious scenario, a user with malicious intent could extract such sensitive information by crafting a specific prompt. If the attacker successfully discerns that the stolen code snippet originates from the training data, they could proceed to pilfer the training dataset or extract further sensitive data from the model.

Determining whether a given code has been employed in the training of a code completion system presents significant challenges, primarily due to the limited insights afforded by black-box access to the model's outputs.
Firstly, as the provided code completion service consistently operates as a black box, users are constrained in their ability to query it, imposing limitations on available information for discerning whether a specific code has indeed contributed to the model's training dataset.
Secondly, the code completion scenario under examination is invariably represented as either sequence generation or a chained sequence of classifications, which further intensifies the complexity of the problem at hand.
From our investigation, existing works on this related problem mainly target neural classification models. It is comparatively straightforward to detect information leakage by scrutinizing the output labels of classification models, while the sequence generation models might obfuscate the information leakage as labels are generated consecutively with dependencies.

\noindentparagraph{\textbf{\textup{Our Solution and Contributions.}}}
To tackle the aforementioned challenges, this paper tailors a membership inference approach (termed \methodnospace) that was originally crafted for classification tasks to a more challenging task of code completion. 
We have reframed our problem as follows: 
\textit{``Given a code sample and black-box access to a target neural code completion model, our objective is to ascertain whether the sample has been used in the model's training dataset.''}
To start with, as the specifics (e.g., architecture and parameters) of the target neural code completion model remain undisclosed to users, we train multiple shadow models to emulate the behavior of the target model. 
Leveraging the outputs of these shadow models, we convert the posterior probability vectors into a set of ranks (\textit{cf.} Section~\ref{sec_membership_classifier}). Subsequently, we develop a neural membership classifier using these rank features to determine whether the provided code sample belongs to the training dataset.
We comprehensively evaluate the effectiveness of this adapted approach across a diverse array of neural code completion models, (i.e., LSTM-based~\cite{hochreiter1997long}, CodeGPT~\cite{lu2021codexglue}, CodeGen~\cite{nijkamp2023codegen}, and StarCoder~\cite{li2023starcoder}).
Experimental results reveal that the LSTM-based and CodeGPT models suffer the membership leakage issue, which can be easily detected by our proposed membership inference approach with an accuracy of 0.842, and 0.730, respectively.

To summarize, the key contributions of this paper are as follows:
\setlist[itemize]{left=0pt}
\begin{itemize}
    \item 
    We conduct an exploratory and systematical research around the legal and ethical issues of existing neural code completion systems.
    
    \item 
    We present a novel membership inference approach (termed \methodnospace) that was originally crafted for classification tasks to a more challenging task of code completion. 
    
    \item 
    We perform comprehensive experiments to measure the efficacy of the membership inference approach against four state-of-the-art neural code completion models (i.e., LSTM-based~\cite{hochreiter1997long}, CodeGPT~\cite{lu2021codexglue}, CodeGen~\cite{nijkamp2023codegen}, and StarCoder~\cite{li2023starcoder}). 
    Experimental results demonstrate the effectiveness of our membership inference approach in determining whether a given code sample has been utilized in the model's training dataset.
\end{itemize}

\section{Motivating Examples}
In this section, we illustrate the motivation of determining whether a given code has been used to train a code completion system by presenting two real-world examples.

\subsection{Example \#1: A Lawsuit against GitHub Copilot}
A user can protect their intellectual property in-licensed source code by discerning whether the source code has been utilized for model training.
We use a real-world lawsuit to illustrate the importance of protecting the intellectual property of source code in current neural code completion models.
GitHub Copilot~\cite{copilot}, a Visual Studio plugin released by Microsoft for code generation in 2022, has attracted increasing attention recently. 
This tool is built on OpenAI's Codex~\cite{codex}, which is trained on billions of publicly available source code snippets and natural-language descriptions, including samples from public repositories on GitHub.
Nonetheless, the open-source code used for model training is always released under a certain license, and the obligations imposed by the license (e.g., preserving accurate attribution of the source code) must be complied with. 
Yet, GitHub Copilot did not pay enough attention to this issue, resulting in many lawsuits 
concerning alleged open-source license copyright violations arising from its use.

\begin{figure*}[!t]
	\centering
	\begin{minipage}{0.5\textwidth}
		\includegraphics[width=\textwidth]{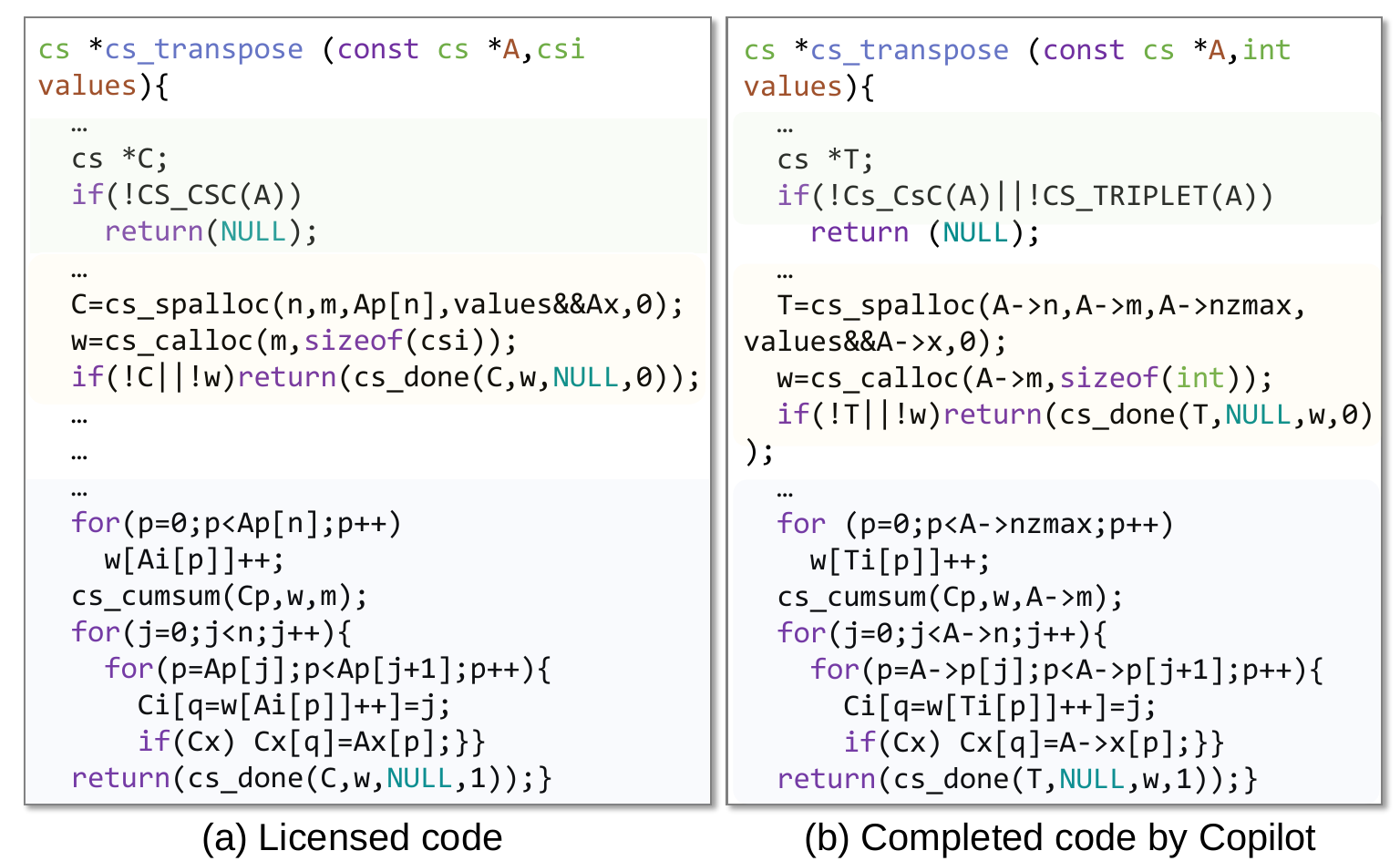}
        \caption{A piece of evidence that Copilot is infringing the copyright of users in code generation.}
		\label{fig_copilot_example}
	\end{minipage}
	\hfill 
	\begin{minipage}{0.46\textwidth}
		\includegraphics[width=\textwidth]{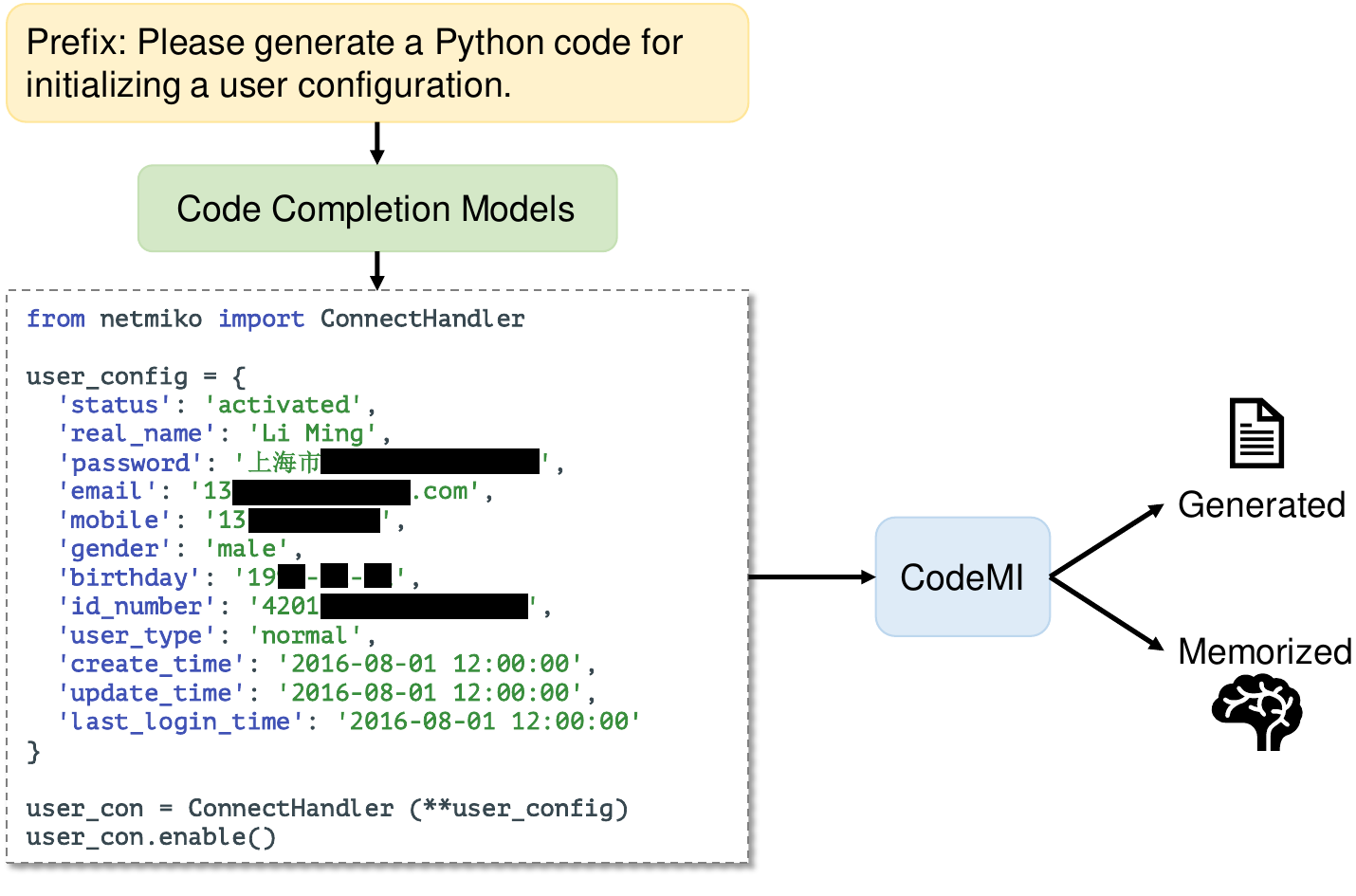}
            \caption{\method can facilitate the steal of private training data that the model memorizes.}
		\label{fig_data_leakage}
	\end{minipage}
\end{figure*}

Figure~\ref{fig_copilot_example} presents a piece of evidence that Copilot is infringing the copyright of users during model training. 
It is evident that Copilot frequently generates substantial code segments, directly replicated from identifiable repositories, without providing proper attribution for the original code's licensing.
To elaborate, consider the prompt ``\textit{sparse matrix transpose, cs\_}'', Copilot autonomously produces a function implementation named \texttt{cs\_transpose} for transposing sparse matrices within the \texttt{CSparse}, a concise sparse matrix package, as shown in the Figure~\ref{fig_copilot_example} (b).
It is apparent that the code generated by Copilot bears a striking resemblance to code originally authored by Tim Davis\footnote{Source: https://twitter.com/DocSparse/status/1581461734665367554?cxt=HHwWhMDRibPEvfIrAAAA}, as evidenced in Figure~\ref{fig_copilot_example} (a).

\subsection{Example \#2: Stealing the Private Training Data}
Recent studies~\cite{niu2023codexleaks,yang2024unveiling} have revealed that existing code language models possess the capability to memorize certain aspects of their training data, thereby potentially generating sensitive information.
This training data may span from personally identifiable information (e.g., emails and social media accounts) to confidential data (e.g., SSNs and medical records) and even secret information (e.g., passwords, access keys, and PINs). Moreover, the training data may also contain data from both publicly available sources and conceivably private user-generated code~\cite{meli2019bad}.
Figure~\ref{fig_data_leakage} presents a real-world case to illustrate the memorization issues that the code completion models (i.e., StarCoder) are suffering.
Given a textual prompt ``\textit{please generate a Python code for
initializing a user configuration.}'', the StarCoder tends to generate a code that contains private information, such as the username, Email, birthday, and password.
In this scenario, if the attackers successfully discern that a code snippet originates from the training data via our proposed \methodnospace, they could proceed to steal the training dataset or extract further sensitive data from the model.
\section{Background}
\label{sec:background}
In this section, we begin by providing an introduction to some essential background concepts, including neural code completion and membership inference.

\subsection{Code Completion}\label{sec2.1}
\begin{wrapfigure}{R}{0.56\textwidth}
\centering
\includegraphics[width=0.56\textwidth]{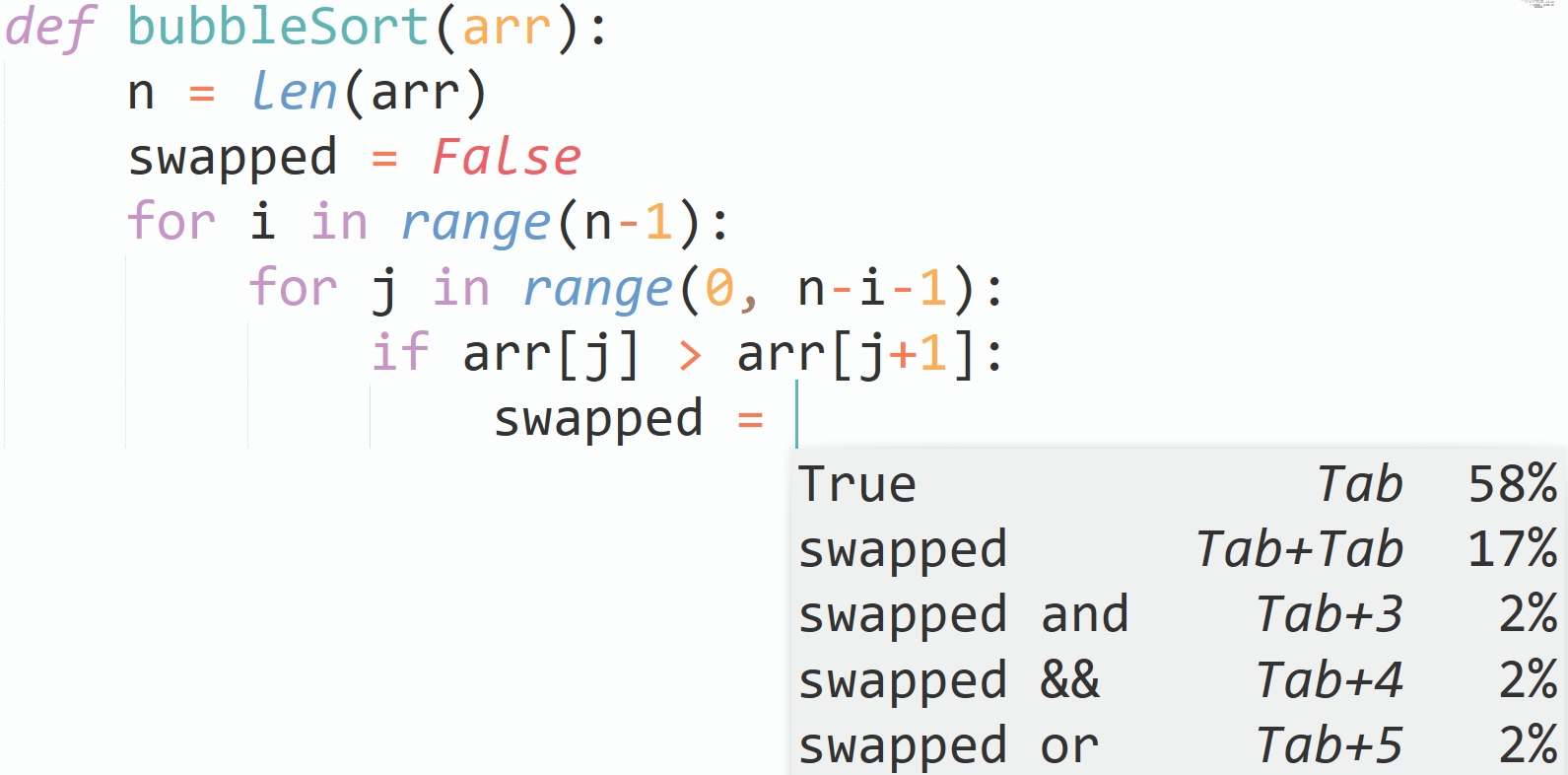}
\caption{An example of code completion by the TabNine plugin.}
\label{fig_code_completion_introduction}
\end{wrapfigure}
The automatic completion of source code, based on the partial code that is being typed, has emerged as a pivotal feature within contemporary IDEs. This feature serves to enhance the productivity of developers engaged in computer programming.
Primarily, the majority of code completion tools are designed to predict the subsequent code elements, such as variable names and APIs, based on the partially typed code. We refer to this particular form of code completion as ``token-level code completion''.
Recently, there has been a surge in the development of tools that aspire to automatically complete entire lines of code or even entire function blocks, referred to as ``line-level'' and ``block-level'' code completion, respectively.
To clarify the scope of this paper, this paper focuses on examining token-level code completion systems.
To illustrate this concept, Figure~\ref{fig_code_completion_introduction} showcases an illustrative example of the prediction of the next code token, as demonstrated by the TabNine plugin~\cite{tabnine}.

\noindentparagraph{\textbf{\textup{Code Completion as a Service.}}}

In practice, the neural code completion models are deployed online to provide services by request.
Only black-box APIs are furnished, enabling users to submit their input queries (i.e., partial code). Consequently, the specifics of the model, including its architecture and parameters, as well as the specifics of its training, remain concealed from users.

\subsection{Neural Language Models}\label{sec2.2}
The fundamental architecture at the core of current neural code completion systems is the language model, the primary objective of which is to predict the probability distribution for generating the next code token, based on the provided partial code.
Let $c$ represent a code snippet, which can be tokenized into a sequence of code tokens, denoted as $\{w_1, w_2, \ldots, w_{|c|}\}$. Furthermore, let $\mathcal{V}$ denote the vocabulary used for mapping each code sample to corresponding indexes.
Given a code sample $c = \{w_1, w_2, \ldots, w_{|c|}\}$, the language model can estimate the joint probability $p(c)$ as a product of conditional probabilities, as follows:
\begin{equation}
    p(c)=\prod_{i=1}^{|\mathbf{c}|} p(w_i|{w_1,w_2,\ldots,w_{i-1}}) \,.
\end{equation}
State-of-the-art neural language models typically calculate this probability based on Recurrent Neural Networks (e.g, LSTM~\cite{hochreiter1997long} and GRU~\cite{chung2014empirical}) as well as Transformers~\cite{vaswani2017attention}.
The language model can undergo training on an extensive code corpus denoted as $\mathcal{D} = \{c_1, c_2, \ldots, c_N\}$, where $N$ represents the corpus's size. Consequently, the loss function employed for the model training is the negative log-likelihood:
\begin{equation}
    \mathcal{L}_{lm}(\theta) = -\sum_{c\in \mathcal{D}}\sum_{i=1}^{|c|} \log p(w_i|{w_1,w_2,\ldots,w_{i-1}};\theta)\,,
\end{equation}
where $\theta$ is the model parameters.
Once the language model is obtained, during the inference stage, we can iteratively generate the subsequent code token by selecting high-probability tokens from the entire vocabulary.

\noindentparagraph{\textbf{\textup{Target Neural Code Completion Models.}}}
In this paper, we select four representative neural code completion models previously proposed in the literature as our focal models for investigation.
It is straightforward to extend this study to other code completion systems.

\noindent
{\textbf{$\triangleright$ LSTM-based~\cite{hochreiter1997long}.}} 
It is natural to model the source code as sequential code tokens like natural-language texts, and then apply an RNN (e.g., LSTM~\cite{hochreiter1997long} and GRU~\cite{chung2014empirical}) to represent the contextual partial code.
In this work, we adopt LSTM to represent  source code, which has also been adopted in Pythia~\cite{svyatkovskiy2019pythia}, a code completion tool for Python.

\noindent
{\textbf{$\triangleright$ CodeGPT~\cite{lu2021codexglue}.}} 
CodeGPT is an autoregressive pre-trained language model designed for code completion, following the architecture of GPT-2~\cite{brown2020language}. 
In both CodeGPT and GPT-2, the cornerstone is the Transformer network~\cite{vaswani2017attention}, an essential module extensively utilized across various NLP tasks.

\noindent
{\textbf{$\triangleright$ CodeGen~\cite{nijkamp2023codegen}.}} 
CodeGen is a code generation model that employs a multi-turn program synthesis approach. It uses autoregressive Transformers to generate code based on user intents and is trained on both programming language and natural language data.

\noindent
{\textbf{$\triangleright$ StarCoder~\cite{li2023starcoder}.}} 
StarCoder is an LLM specifically crafted for code, having undergone pre-training on an impressive corpus of 1 trillion code tokens drawn from The Stack~\cite{Kocetkov2022TheStack}. 
It is also fine-tuned on a substantial dataset of 35 billion Python tokens, containing 15.5B parameters.

\subsection{Membership Inference}
\begin{figure}[!t]
\centering
\includegraphics[width=0.9\textwidth]{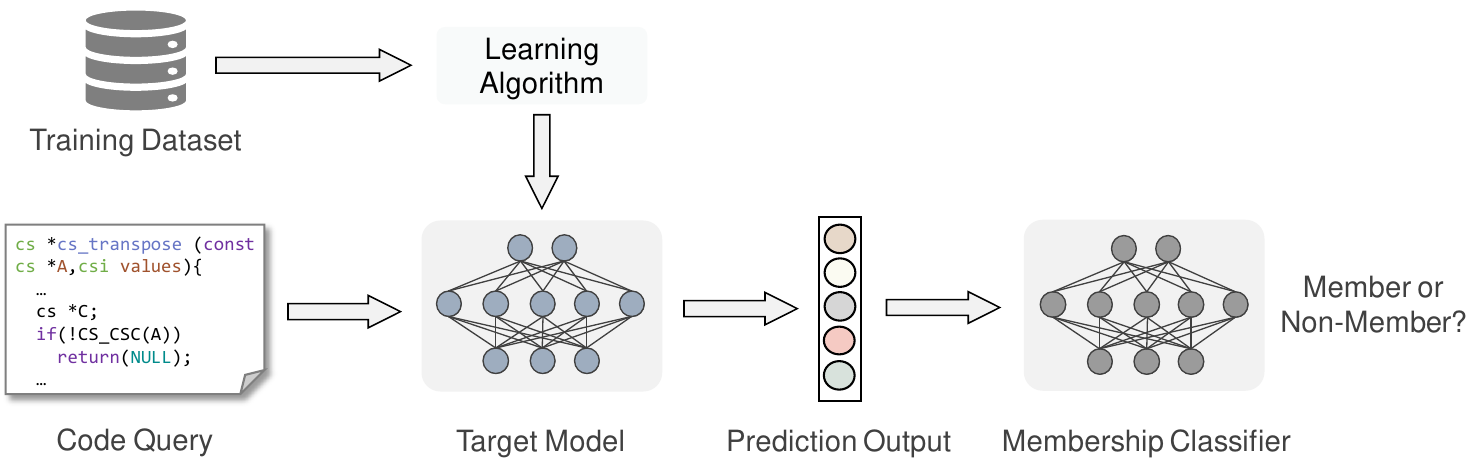}
\caption{The pipeline of membership inference.}
\label{fig_membership_inference_pipeline}
\end{figure}

Membership inference\footnote{In some literature in security, it is also termed \textit{membership inference attack}.} 
is a task focused on determining whether a given data sample has been employed in the training of a machine learning model~\cite{hu2022membershipsurvey}.
Figure~\ref{fig_membership_inference_pipeline} shows the pipeline of membership inference.
Let $f(c;\theta^*)$ denote the target code completion model that is trained on a dataset $\mathcal{D}_{train}=\{c_1, c_2,\ldots, c_N\}$, where $\theta^*$ is the model parameters.
The fundamental concept behind membership inference is to leverage discrepancies in the behavioral patterns, specifically the predicted outputs, of the target model when presented with member and non-member data.
In essence, the model is anticipated to exhibit superior performance on member data in comparison to non-member data, owing to its exposure to and memorization of the member data during the training phase.
When presented with a novel data sample $x = \{w_1, w_2, \ldots, w_{|x|}\}$, the target code completion model $f(x; \theta^*)$ processes it as input, producing a sequence of probability vectors denoted as $h_x = \{p(w_2 | w_1), p(w_3 | w_1, w_2), \ldots, p(\texttt{[EOS]} | w_1, w_2, \ldots, w_{|x|})\}$. Here, \texttt{[EOS]} denotes the end of the sequence.
Each probability vector indicates the likelihood of generating the next code token.
Subsequently, a membership classifier $g(h_x; \eta^{*})$ can be trained utilizing these probability vectors as input features:
\begin{equation}
    g(h_x; \eta^{*}) \rightarrow \{0, 1\}\,,
\end{equation}
where 1 and 0 denote whether the query code sample $x$ belongs to or does not belong to the training dataset of $f(c,\theta^*)$, respectively.

Based on the knowledge accessible to users, membership inference approaches can be classified into two distinct categories: \textit{black-box} and \textit{white-box} membership inference.
In the black-box setting, neither the dataset (including its underlying data distribution) nor the model architecture are accessible. When dealing with an arbitrary input $x$, the user\footnote{In some literature in security, the user is also termed an adversary or attacker from the perspective of attacking.} is restricted to black-box access to the target model, from which they can extract a sequence of prediction vectors $h_x$.
Conversely, in the white-box setting, the user possesses comprehensive knowledge about the target model, including both its architecture and parameters. As such, the black-box setting, with its limited knowledge availability, is considered a more realistic scenario than the white-box setting.

\noindentparagraph{\textbf{\textup{Key Aspects and Challenges.}}}
\label{sec3.4}
It is worth highlighting several key aspects of this study as follows.
\begin{itemize}
    \item \textbf{{Black-Box Target Models.}} In this paper, we aim to study a more realistic scenario of code completion under the black-box setting, where the user is only allowed to access the model and obtain the predicted results.
    Note that, in our experimental evaluation, even though we have implemented two code completion models (i.e., LSTM-based and CodeGPT) as targets, we still consider them as black boxes.
    \item \textbf{{Partial Output of Target Models.}} 
    In the actual context of code completion, the generated completion output is 
    often partial. This means that the list of predicted tokens, ordered by their likelihood, has been intentionally truncated to fit within the constraints of the user interface.
    As illustrated in Figure~\ref{fig_code_completion_introduction}, the predicted output is presented as a concise list of 5 tokens.
    \item \textbf{{Limited Times of Queries.}} 
    In a real-world scenario, users engage with the code completion service through a pay-as-you-go subscription. As an illustration, TabNine's free account exclusively accommodates short code completions, typically spanning 2 to 3 words. In other words, each invocation of the code completion API carries an associated fee. This motivates us to carry out our membership inference study with a limited number of queries. We investigate the impact of query quantity through experiments conducted in Section~\ref{sec4.6}.
\end{itemize}

\section{Our Approach: \method}
\label{sec3}

\begin{figure*}[!t]
\centering
\includegraphics[width=0.92\textwidth]{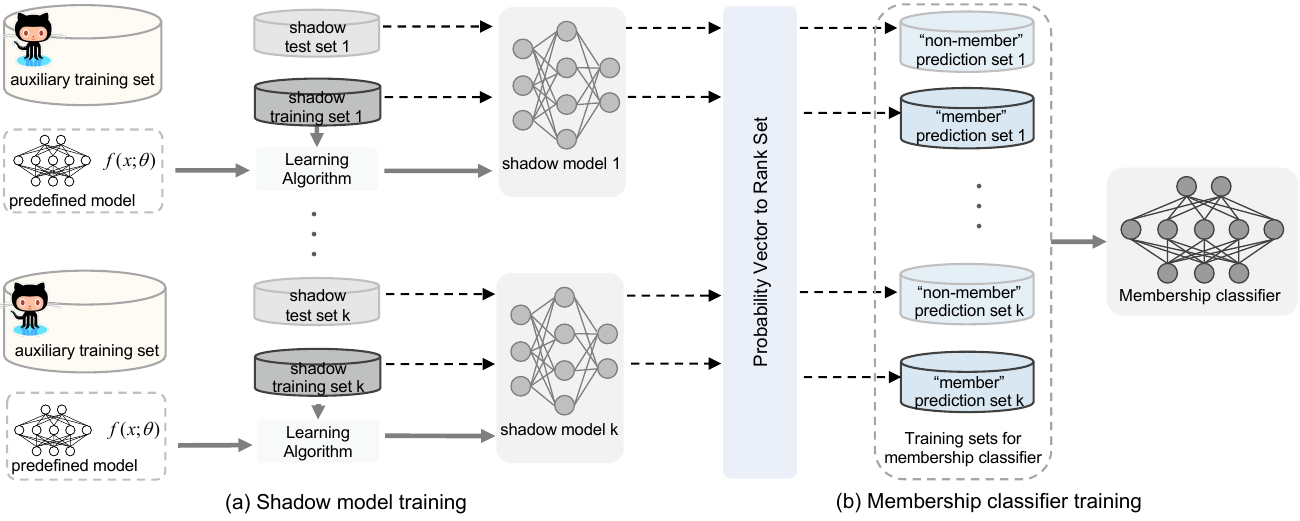}
\caption{An overview of \method.
}
\label{fig_approach}
\end{figure*}

In this section, we introduce a membership inference approach, referred to as \method, which is designed to  determine whether a given code sample has been utilized in the training of a target code completion model.
Our \method adheres to the shadow training paradigm, initially introduced in~\cite{shokri2017membership} for membership inference, and it has inspired numerous related studies~\cite{hu2022membership, hu2022membershipsurvey}. Figure~\ref{fig_approach} provides an overview of our proposed \methodnospace, grounded in the concept of shadow model training.
Specifically, our \method is mainly composed of three modules: \textit{(1) Shadow model training.} 
Since the specifics of the target model, e.g., its architecture and parameters, remain concealed from users, we address this by training multiple shadow models to mimic the behavior of the target model.
\textit{(2) Membership classifier training.} 
In this module, we start by constructing a dataset for training the membership classifier. For each shadow model that has been trained, we employ its respective training and test datasets as input queries to generate corresponding output vectors. 
The predicted output vectors for instances within the training dataset are labeled as ``member'', whereas the predicted output vectors for instances in the test dataset are labeled as ``non-member''.
Subsequently, these labeled predicted vectors serve as the training dataset for the binary membership classifier.
\textit{(3) Membership inference.} 
Leveraging the membership classifier we have obtained, we gain the capability to perform membership inference for a given code sample. Specifically, the code sample is first input into the target model as a query, and subsequently, the membership classifier determines the class label based on the predicted output vector.

\subsection{Shadow Models}
\label{sec3.1}
Following~\cite{shokri2017membership}, we start by training several shadow models with the objective of mimicing the behavior of the target model.
Let $\{f'(\theta_1), f'(\theta_2), \ldots, f'(\theta_k)\}$ denote a set of $k$ shadow models.
Since many code corpora used for code model training are collected from open-source communities such as GitHub, we have not completely avoided the overlap between the training datasets of the shadow model and the target model.
Based on this assumption, in this paper, we use the partial Py150 dataset as an auxiliary dataset for shadow model training.

Even though the neural architecture of target models is unknown, we can guess the model architecture
and use it to define the architecture of shadow models.
In this paper, we have designed two distinct architectures for shadow models: LSTM-based~\cite{hochreiter1997long} and Transformer~\cite{lu2021codexglue}.
Both models are trained within the framework of language modeling, as described in Section~\ref{sec2.2}.
The specific training data construction of shadow models and implementation process are described in Section~\ref{sec5.2}.

\subsection{Membership Classifier}\label{sec_membership_classifier}
\subsubsection{Training Data Construction}
Initially, we create a dataset dedicated to training the membership classifier. Specifically, for each trained shadow model, we label the predicted output vectors of instances within the training dataset $\mathcal{D}_{shadow}^{train}$ as ``member''. Subsequently, we employ the test dataset $\mathcal{D}_{shadow}^{test}$ as our input queries and label the corresponding predicted output vectors as ``non-member''.

\subsubsection{Probability Vectors to Rank Set}
\begin{figure}[!t]
\centering
\includegraphics[width=0.7\textwidth]{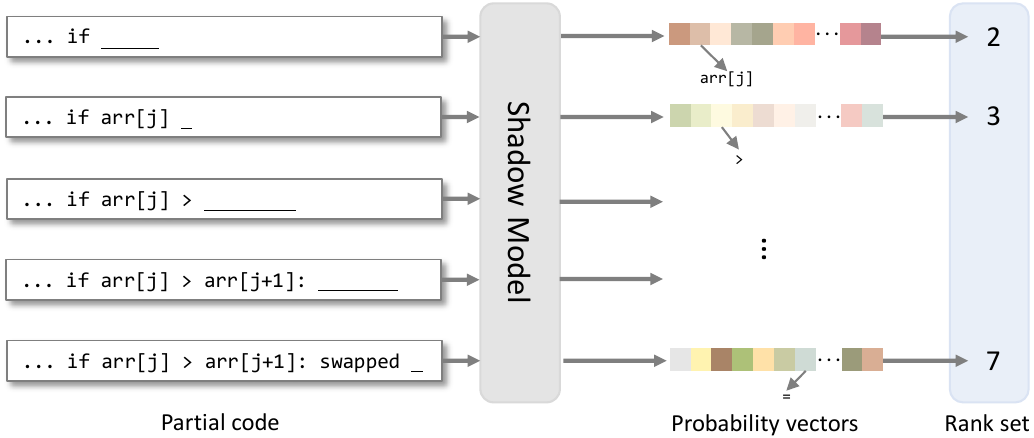}
\caption{A toy example to illustrate the process of converting probability vectors to rank set.}
\label{fig_probability_to_vector}
\end{figure}

Typically, the predicted output vector takes the form of a probability distribution spanning the entire training vocabulary, denoted as $\mathcal{V}$. In other words, it represents a probability vector with $|\mathcal{V}|$ dimensions.
For example, given a code sample $x=\{w_1, w_2,\ldots,w_{|c|}\}$, the user queries the shadow model at each time step and obtains the probability vector for generating the next token $p(y_{i+1}|w_i)=f(w_i,\theta^*)$.
However, when using the code completion services in real-world scenarios, in most cases the user can only obtain a list of candidate tokens ranked from high to low according to their probability, but can't obtain the model output logits corresponding to each candidate token, i.e., probability values. This inspires us to 
utilize the ranks of ground-truth words within the output distributions as features to infer the membership of a given code sample, rather than using the raw probability values.
Figure~\ref{fig_probability_to_vector} gives an illustration of converting probability vectors to a rank set.
Let $h_x^{(p)}=\{p(w_2|w_1), p(w_3|w_1, w_2),\ldots, p(\texttt{[EOS]}|{w_1,w_2,\ldots,w_{|x|}})\}$ denote a sequence of probability vectors, where $p(y_{t+1}|w_t)$ denotes the probability vector of generating code token, where the ground-truth is $\widehat{w}_{t+1}$. We denote the rank of probability of generating $\widehat{w}_{t+1}$ as $R(\widehat{w}_{t+1}|w_{t})$.
Finally, we can convert a sequence of probability vectors $h_x^{p}$ into a rank set $h_x^{r}=\{R(\widehat{w}_{2}|w_{1}), R(\widehat{w}_{3}|w_{2}), \ldots, R(\widehat{w}_{\texttt{[EOS]}}|w_{|x|})\}$.

Taking a real code completion case as an example, where the typed partial code is ``\texttt{... if \underline{arr[j]} \underline{>} \underline{arr[j+1]}: \underline{swapped} \underline{=}}''.
Assuming that the code completion model has been triggered 5 times, i.e., in predicting \texttt{arr[j]}, \texttt{>}, \texttt{arr[j+1]}, \texttt{swapped}, and \texttt{=} (marked underline).
Let $\{p(\texttt{arr[j]}|ctx), p(\texttt{>}|ctx), p(\texttt{arr[j]+1}|ctx), p(\texttt{swapped}|ctx),  p(\texttt{=}|ctx)\}$ be the sequence of five probability vectors for tokens ``\texttt{arr[j]}'', ``\texttt{>}'', ``\texttt{arr[j+1]}'', ``\texttt{swapped}'', and ``\texttt{=}'', respectively.
The probabilities of these ground-truth tokens are ranked in $2$, $3$, $6$, $10$, $7$, respectively.
Therefore, the rank set converted from the output probability vectors for the given code sample is $h_r=\{2, 3, 6, 10, 7\}$. 
Rank 2 means that the word ranks as the second most probable prediction across the entire vocabulary. Subsequently, after collecting ranks for all predictions, the user constructs a histogram for $h^{(r)}$ utilizing a fixed number of bins.
Finally, we obtain a paired dataset of the predicted rank set and its corresponding membership labels. 
The labeled rank set is used as a training dataset $\mathcal{D}_{mc}$ to train the binary membership classifier.

\subsubsection{Neural Model Training}
We train a binary membership classifier based on its training dataset $\mathcal{D}_{mc} = \{(h_{x_1}^{(r)}, y_1), (h_{x_2}^{(r)}, y_2), \ldots, (h_{x_m}^{(r)}, y_m)\}$. Following the previous works~\cite{shokri2017membership,salem2019ml}, a Multi-Layer
Perceptron (MLP) is used as the binary classifier.
We employ the cross-entropy loss function as our optimization objective, outlined as follows.
\begin{equation}
\begin{split}
    \mathcal{L}_{mc}(\beta) & = -\sum_{i=1}^{m} [ p(y_i=1|h_{x_i}^{(r)};\beta)\log p(y_i=1|h_{x_i}^{(r)};\beta) \\
    & + p(y_i=0|h_{x_i}^{(r)};\beta)\log p(y_i=0|h_{x_i}^{(r)};\beta) ] \,,
\end{split}
\end{equation}
where $p(y_i=1|h_{x_i}^{(r)};\beta)$ and $p(y_i=0|h_{x_i}^{(r)};\beta)$ denotes the probability of classifying the input as member and non-member, respectively. $\beta$ denotes the parameters of the membership classifier.

\subsection{Membership Inference}
During the inference phase, when provided with a code sample denoted as $x$, the user can ascertain its membership by submitting it to the target model $f(x; \theta^*)$ and subsequently inputting the resulting prediction into the membership classifier $g(h_x^{(r)};\beta)$.
In particular, following the query of $f(x, \theta^*)$, we obtain the corresponding prediction vectors $h_x^{(p)}$, which are then converted into a rank set denoted as $h_x^{(r)}$.
Ultimately, the membership classifier accepts the transformed rank set $h_x^{(r)}$ as its input and makes predictions regarding whether the code sample has been included in the training dataset of the target model.
As previously mentioned, the target model may impose limitations on the number of queries it can make. 
In such situations, two strategies emerge: prioritizing queries with the lowest frequency counts or opting for random selection of partial queries. This approach resonates with the notion that sequences containing uncommon terms are inherently more discernible.

\section{Experimental Setup}
This section sets up the experiments, aiming to address the following four research questions:

\begin{itemize}
    \item \textbf{RQ1:} How effective is membership inference in determining if a code sample was utilized to train a neural code completion model?
    \item \textbf{RQ2:} How does the number of the shadow model impact our membership inference approach?
    \item \textbf{RQ3:} What is the impact of the output size of the target model on our membership inference approach?
    \item \textbf{RQ4:} How does the number of queries affect our membership inference approach?
\end{itemize}

\subsection{Target Model Implementations}
\label{sec5.1}
We independently re-implemented the target code completion models to rigorously assess the efficacy of \method during the testing phase, but we did not use any a priori knowledge about the target models throughout the experimental process of \method (including the shadow models training and the membership inference phases), which always followed the black-box setup. This deliberate approach ensures the integrity and impartiality of our evaluation. This is also the reason why we do not use Copilot mentioned in Section~\ref{sec_introduction} as the target model. Even if the inference result of membership  for the Copilot model data turn out to be consistent with the facts, the lack of insight into its training data  undermines the ability to confidently validate \method.

We implement the four target models in the following manner.

\smallskip
\noindent{\textbf{$\triangleright$  LSTM-based.}}
In line with the approach in~\cite{kim2021code}, we employ an LSTM-based model for code completion. Our training data source is Py150 
which comprises 150,000 Python source code files gathered from GitHub repositorie. 
For model training purposes, we partition the dataset into two subsets: 24,000 files for the member dataset and another 24,000 for the non-member dataset.
This model is constructed using a single LSTM layer, equipped with a vocabulary size encompassing 100,002 distinct code tokens. The embedding dimension is set at 300, and model training is conducted using the Adam optimizer with a learning rate of 1e-3, a batch size of 4, and a maximum of 20 epochs. To avoid overfitting, we employ a dropout mechanism with a ratio of 0.5.

\smallskip
\noindent{\textbf{$\triangleright$  CodeGPT.}}
We utilize the pre-trained model \texttt{CodeGPT-small-py}, which is introduced in the CodeXGLUE~\cite{lu2021codexglue}.
To tokenize the source code, we employ Byte-Pair Encoding (BPE) and configure the vocabulary size to 50,234. Fine-tuning of CodeGPT is carried out on the Py150 dataset, spanning 5 epochs and utilizing a batch size of 4. Adam optimization is employed with a learning rate set to 8e-5.
It is essential to note that the training data employed for fine-tuning this model originated from the Py150 dataset and serves as member data. To ensure that the non-member dataset used does not overlap with the training dataset of the target model, we leverage a tool described in~\cite{xu2022systematic}. This tool allows us to crawl approximately 100,000 deduplicated Python files from GitHub repositories created no earlier than January 1, 2023, and with no fewer than 20 stars. This selection criteria ensures both the freshness of the code samples and their high quality.
For the sake of clarity in subsequent discussions, we shall refer to this dataset as Github-Python. It's important to highlight that a subset of Github-Python was chosen as CodeGPT's non-member dataset.

\smallskip
\noindent
{\textbf{$\triangleright$  CodeGen \& StarCoder.}} 
We conduct direct inference using the pre-trained code language models \texttt{CodeGen-2B-mono} and StarCoder with 15.5B parameter as our target models.
Following the training guidelines provided by CodeGen and StarCoder, respectively, we select a portion of the training data from the CodeSearchNet dataset~\cite{husain2019codesearchnet} and The Stack~\cite{Kocetkov2022TheStack}, to form the member dataset for CodeGen and StarCoder, respectively. Additionally, we curate a subset of Github-Python to serve as non-member dataset for CodeGen and StarCoder, respectively.

\subsection{Shadow Model Implementations}
\label{sec5.2}
In practical settings characterized by limited access to target models through black-box interactions, identifying the precise architecture of these models poses a significant challenge. To address this issue, we utilize two distinct models as proxies: an LSTM-based model and a Transformer-based CodeGPT model, which we refer to as our shadow models. The purpose of these shadow models is to mirror two different network structures, providing insights into the underlying architecture of the target models.

The implementation process for these shadow models mirrors that of our target models, as discussed in Section~\ref{sec5.1}. However, it is important to note that the selection of training and testing datasets for the shadow models is performed independently of the datasets used for the target models. 
For LSTM-based shadow models, we use 48,000 code files from Py150 (which do not overlap with the dataset of the target LSTM-based model), 24,000 files for the member dataset and another 24,000 for the non-member dataset. For CodeGPT shadow models, we still use datasets from the same data source as the target CodeGPT model but not overlapping as member and non-member datasets, respectively. In conclusion, we have tried to ensure that the member and non-member datasets of the shadow models do not overlap with the member and non-member datasets of the target models, respectively, so as to ensure our black-box access to the target models.
\subsection{Baselines}
We compare \method against several traditional metrics-based approaches.  \citet{salem2019MLLeaks} implement the membership inference attack using unsupervised binary classification. Specifically, they extract the \textbf{highest posterior} and assess whether this maximum exceeds a predetermined threshold. This approach relies on the idea that the model exhibits greater confidence when processing member data.
Following this approach, we calculate the average maximum prediction confidence per token generated by the target model for both member and non-member data. In our study, where we evaluate membership inference using a balanced dataset, we refrain from setting a fixed threshold. Instead, we designate the top 50\% of code samples with the highest posterior as members, while categorizing the remaining 50\% of code samples as non-members.

Additionally, \citet{carlini2021extracting} delve into data extraction attacks on language models. Their approach involves generating numerous samples from a language model and subsequently utilizing \textbf{perplexity} as a metric to discern whether a sample is part of the training data. Perplexity measures the model's ability to predict a given sample, with lower perplexity scores suggesting a higher likelihood of the sample's inclusion in the training data.
Following a similar strategy, we employ perplexity as an additional metric for predicting a sample's membership status. Analogous to our previous approach, we identify the top 50\% of samples with the lowest perplexity as members, while the remaining samples are categorized as non-members.

\subsection{Evaluation Metrics}
\label{sec5.4}
To evaluate the efficacy of our proposed membership inference approach, we employ four standard metrics commonly used for assessing classification models: Accuracy, AUC, Precision, and Recall.

\smallskip
\noindent{\textbf{$\triangleright$ Accuracy}} is defined as the proportion of correctly predicted samples. This metric is computed by dividing the count of accurately predicted samples by the total number of predicted samples.
Accuracy intuitively reflects the performance of the classification model.

\smallskip
\noindent{\textbf{$\triangleright$ AUC (Area under the ROC Curve).}} 
The performance of a classification model at various classification thresholds is illustrated on a graph known as a ROC curve (receiver operating characteristic curve).
In this curve, the horizontal axis represents the false positive rate (FPR), while the vertical axis represents the true positive rate (TPR).
As a numerical value, the classifier with a larger AUC is better. The AUC metric provides a visual representation of the model's performance.

\smallskip
\noindent{\textbf{$\triangleright$ Precision}} 
is measured by dividing the proportion of positive samples that were correctly identified by the total number of samples categorized as positive (either properly or wrongly).
It assesses how precisely the model classifies a sample as positive.

\smallskip
\noindent{\textbf{$\triangleright$ Recall}} 
simply considers the classification of the positive samples. 
It is determined as the proportion of positively categorized positive samples to all positively classed samples.
It evaluates how well the model can identify positive samples.

\section{Results and Analysis}
In this section, we answer each RQ by analyzing the experimental results. Additionally, we undertake a case study to further understand the performance of our proposed approach. 
\subsection{Performance of Membership Inference (RQ1)}

\begin{table*}[t!]
	\centering
	\caption{Performance of \methodnospace, with various architectures for shadow models and two metrics-based baseline approaches. }
	\label{table1}
        \setlength{\tabcolsep}{6pt} 
        \resizebox{\columnwidth}{!}{
    	\begin{tabular}{l|cccc|cccc|cccc|cccc}
                \thickhline
                &\multicolumn{4}{c|}{\textbf{Shadow$_{LSTM-based}$}} & \multicolumn{4}{c|}{\textbf{Shadow$_{Transformer}$}}& \multicolumn{4}{c|}{\textbf{Highest Posterior}}& \multicolumn{4}{c}{\textbf{Perplexity}} \\
    		\cline{2-17}
    	       &  \textbf{Acc.}   &\textbf{AUC} &\textbf{P}	&\textbf{R}  &  \textbf{Acc.}	&\textbf{AUC}&\textbf{P}	&\textbf{R} &  \textbf{Acc.}	&\textbf{AUC}&\textbf{P}	&\textbf{R} &  \textbf{Acc.}    &\textbf{AUC}	&\textbf{P}	&\textbf{R} \\
            \hline
                LSTM 	&\cellcolor{lightblue}0.842	&\cellcolor{lightblue}0.902	&\cellcolor{lightblue}0.774	&\cellcolor{lightblue}0.962 &0.838	&0.853	&0.772 &0.958	&0.652 &0.649 &0.652  & 0.687  	&0.728 &0.776 &0.730 &0.776 \\
                CodeGPT 	&0.688	&0.725	&0.731 &0.595 &\cellcolor{lightblue}0.730	&\cellcolor{lightblue}0.804	&\cellcolor{lightblue}0.743	&\cellcolor{lightblue}0.703 	&0.617	&0.649	&0.617 &0.617 &0.672	&0.717	&0.672 &0.672\\
                CodeGen 	 &0.512	&0.435	&0.507 &0.967&0.535	&0.656	&0.519	&0.980&0.584	&0.625	&0.584 &0.584	&0.612	&0.661	&0.612 &0.612\\
                StarCoder 	&0.582	&0.590	&0.561 &0.745 &0.548	&0.584	&0.536	&0.723	&0.412	&0.390	&0.412 &0.412	&0.407	&0.388	&0.407 &0.407\\
                \thickhline
    	\end{tabular}}
\end{table*}

Table~\ref{table1} presents the performance of \method across various architectures of shadow models and target models. In this table, we also highlight in \textcolor{blue}{blue} the results obtained under the conditions where both the architecture of the shadow model and that of the target model are identical. From the table, it is evident that \methodnospace, which is based on shadow model training, outperforms two metrics-based baseline methods. 
In our investigation, which encompasses four neural code completion models, we explore two types of shadow models: LSTM-based~\cite{hochreiter1997long} and Transformer~\cite{vaswani2017attention}, based on our underlying assumptions. Both of these shadow models are implemented in accordance with the specifications of the target models, as described in Sections~\ref{sec5.1} and~\ref{sec5.2}. It is worth noting that, in practice, users typically do not possess knowledge of the architecture of the target code completion models. 
The results from the table reveal that our proposed \method can effectively determine whether a given code sample was part of the training data for the LSTM-based code completion model and the CodeGPT model. 
For instance, when employing LSTM-based shadow models and targeting LSTM-based code completion model, \method achieves impressive membership inference performance metrics, including an accuracy of 0.842, an AUC of 0.902, a precision of 0.774, and a recall of 0.962.
Similarly, when employing the Transformer-based shadow models and targeting CodeGPT's  Code Completion model, \method  achieves  good performance metrics, including an accuracy of 0.730, an AUC of 0.804, a precision of 0.743, and a recall of 0.703.
However, the performance drops notably when the target code completion model is either CodeGen or StarCoder, suggesting that there is substantial room for further improvement. 
We attribute this to the fact that these LLMs of code exhibit both a superior capacity for generation and consistently high performance, which leads to their ability to make correct predictions with high confidence even for  both member and non-member data, so membership inference performs poorly on these models.
Further discussion will be elaborated upon in Section~\ref{sed_discussion}.

\begin{tcolorbox}
	\textbf{Answer to RQ1.} 
Our proposed \method demonstrates its effectiveness in discerning whether a given code sample has been employed in training both LSTM-based code completion models and CodeGPT models. Notably, a performance drop is observed when applied to the CodeGen or StarCoder models, suggesting opportunities for further enhancement.
\end{tcolorbox}

\begin{figure}[!t]
\centering
    \begin{subfigure}[b][][c]{.24\textwidth}
	\centering
        \includegraphics[width=\linewidth]{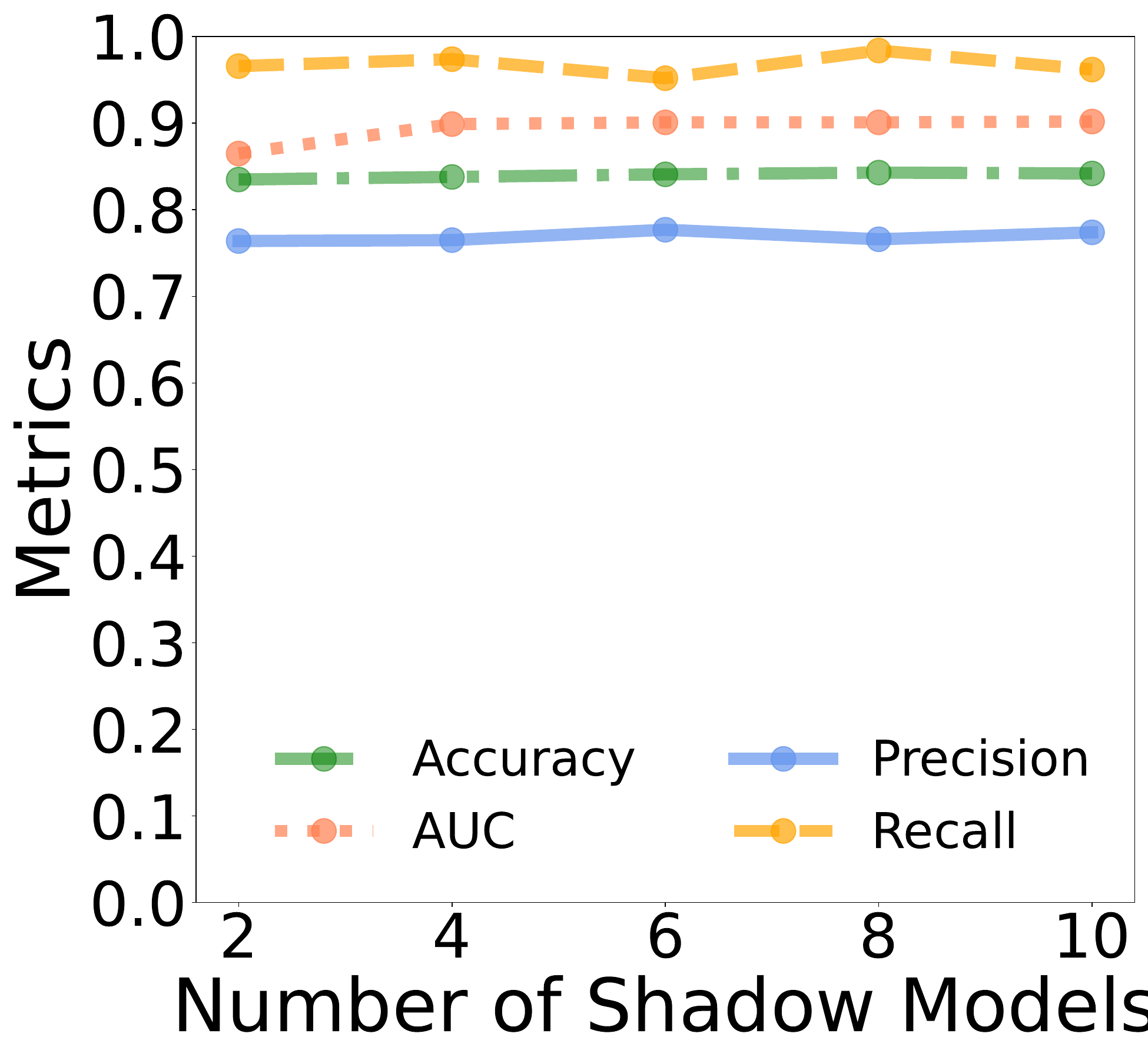}
	\caption{LSTM-based}
	\label{fig_trend_a}
    \end{subfigure}
    \begin{subfigure}[b][][c]{.24\textwidth}
	\centering
        \includegraphics[width=\linewidth]{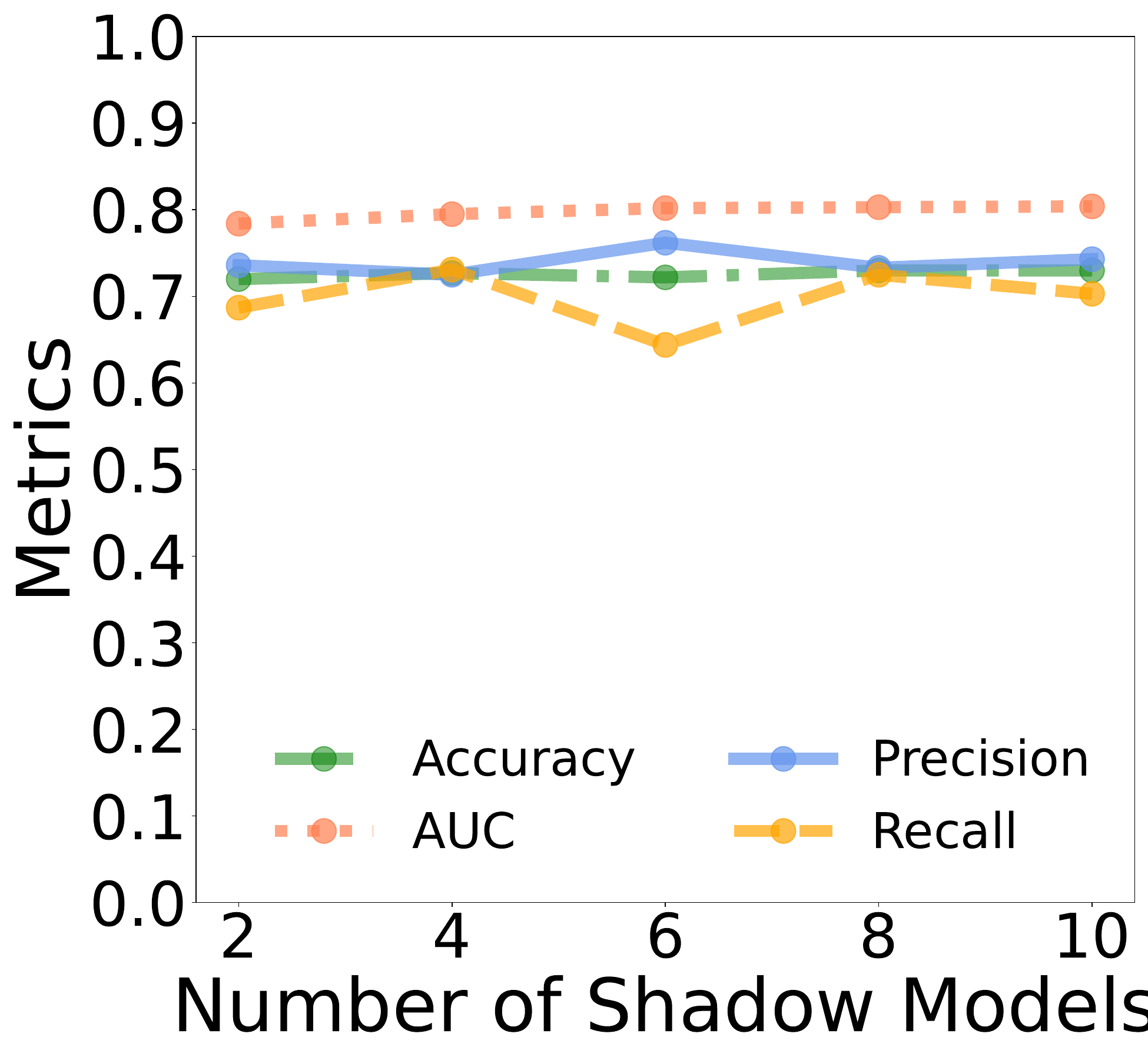}
	\caption{CodeGPT}
	\label{fig_trend_b}
    \end{subfigure}
    \begin{subfigure}[b][][c]{.24\textwidth}
	\centering
        \includegraphics[width=\linewidth]{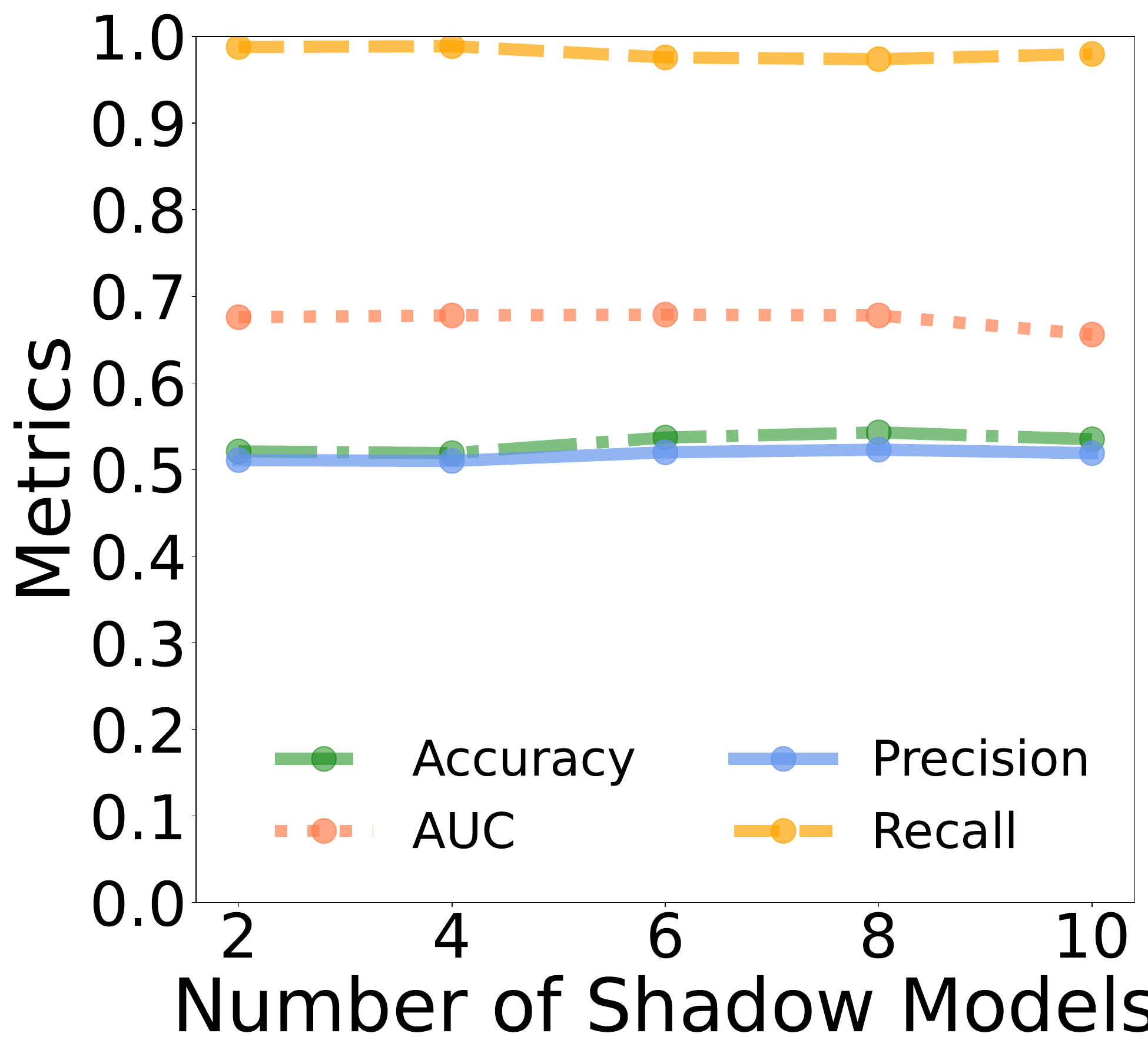}
	\caption{CodeGen}
	\label{fig_trend_b}
    \end{subfigure}
    \begin{subfigure}[b][][c]{.24\textwidth}
	\centering
        \includegraphics[width=\linewidth]{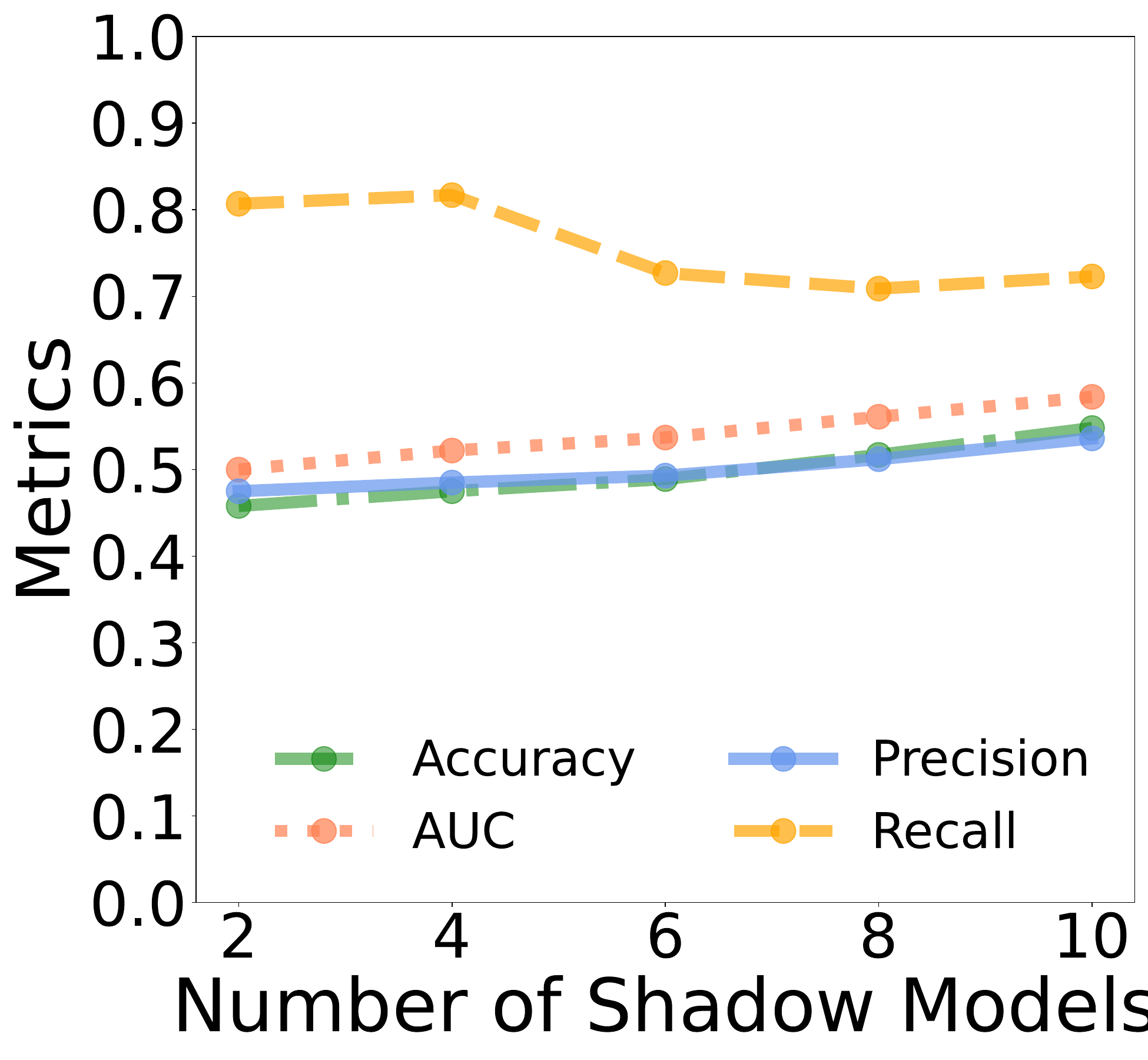}
	\caption{StarCoder}
	\label{fig_trend_b}
    \end{subfigure}
    \caption{The impact of changes in the number of shadow models on membership inference.} 
\label{fig_7}
\end{figure}
\subsection{Impact of the Number of Shadow Models (RQ2)}

In our proposed \methodnospace,
we have devised several shadow models to mimic the behavior of the target code completion model. 
Here, we investigate the influence of varying the number of shadow models on the efficacy of membership inference, as depicted in Figure~\ref{fig_7}.
From this figure, we can observe a consistent trend across shadow models spanning from 2 to 10 (incrementing by 2), wherein there are no discernible fluctuations in Accuracy, AUC, Precision, or Recall.
This suggests that augmenting the quantity of shadow models may not consistently enhance membership inference performance. This observation resonates with earlier findings~\cite{salem2019MLLeaks}. The rationale behind this may lie in the relatively straightforward architecture of our membership classifiers; the training data from multiple shadow models suffices to refine the performance of the membership classifier.
\begin{tcolorbox}
	\textbf{Answer to RQ2.} 
 Our membership inference method doesn't require much shadow models to achieve a comparatively good performance.
\end{tcolorbox}
\begin{figure}[!t]
\centering
    \begin{subfigure}[b][][c]{.24\textwidth}
	\centering
        \includegraphics[width=\linewidth]{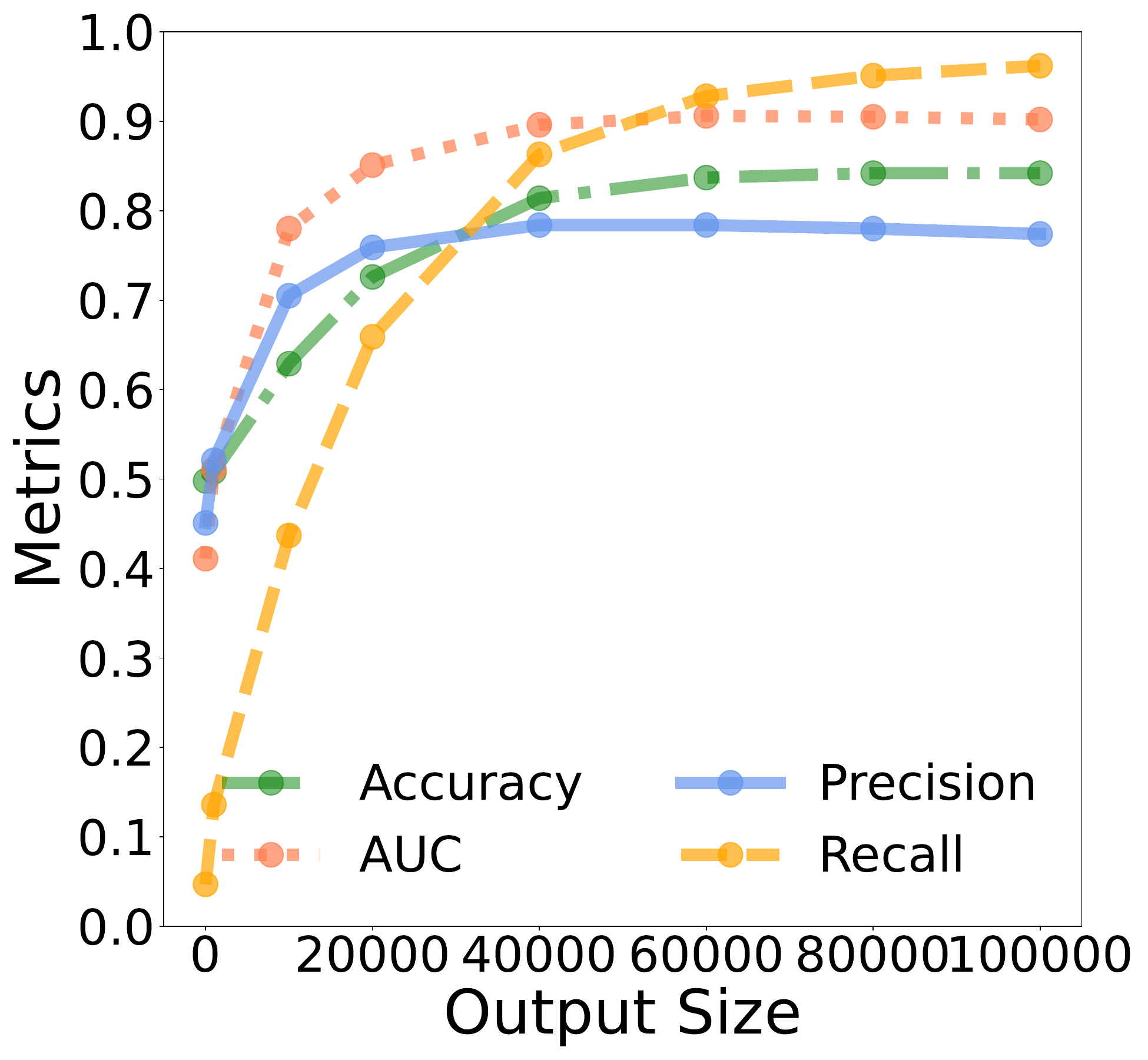}
	\caption{LSTM-based}
	\label{fig_trend_a}
    \end{subfigure}
    \begin{subfigure}[b][][c]{.24\textwidth}
	\centering
        \includegraphics[width=\linewidth]{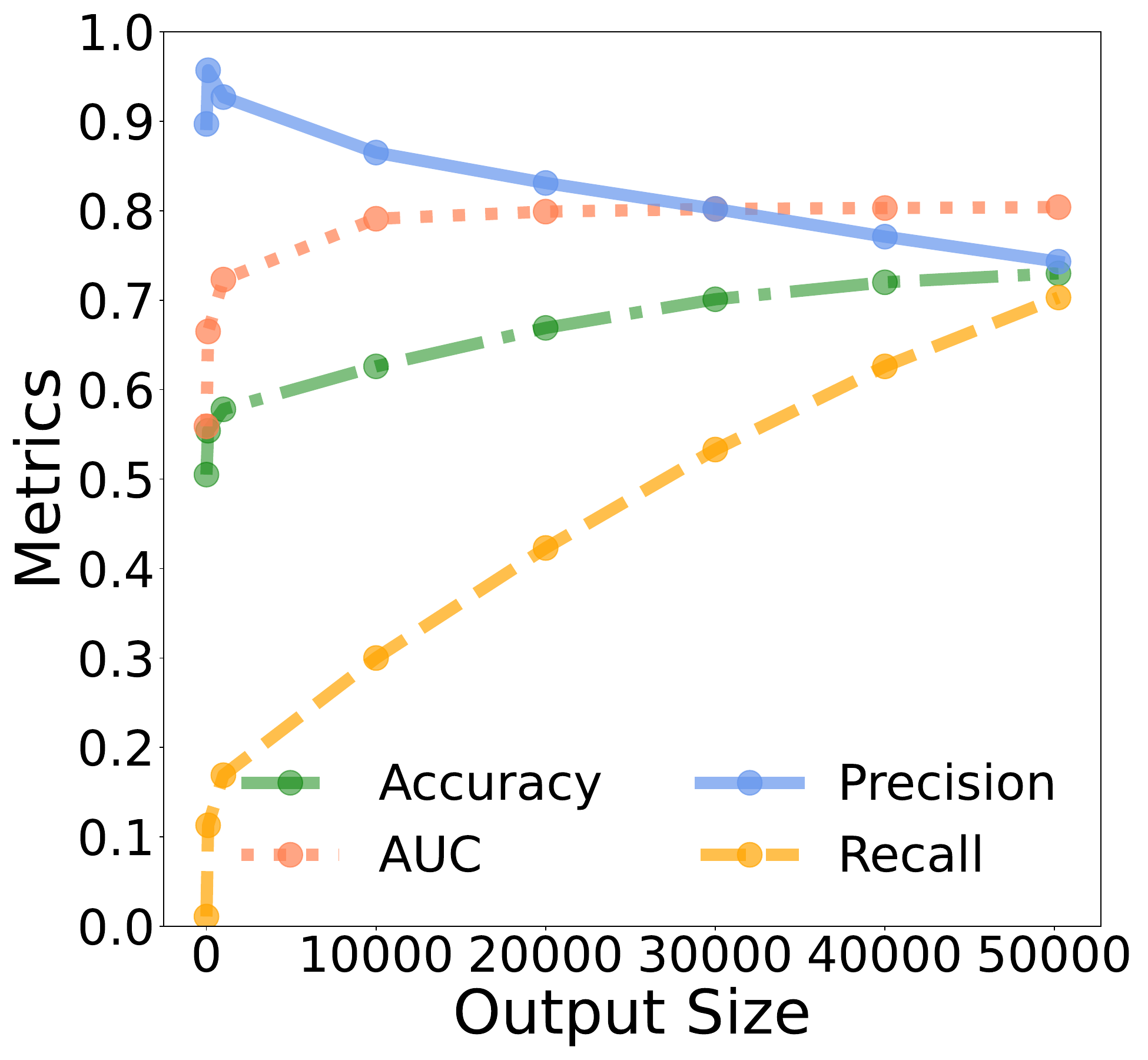}
	\caption{CodeGPT}
	\label{fig_trend_b}
    \end{subfigure}
    \begin{subfigure}[b][][c]{.24\textwidth}
	\centering
        \includegraphics[width=\linewidth]{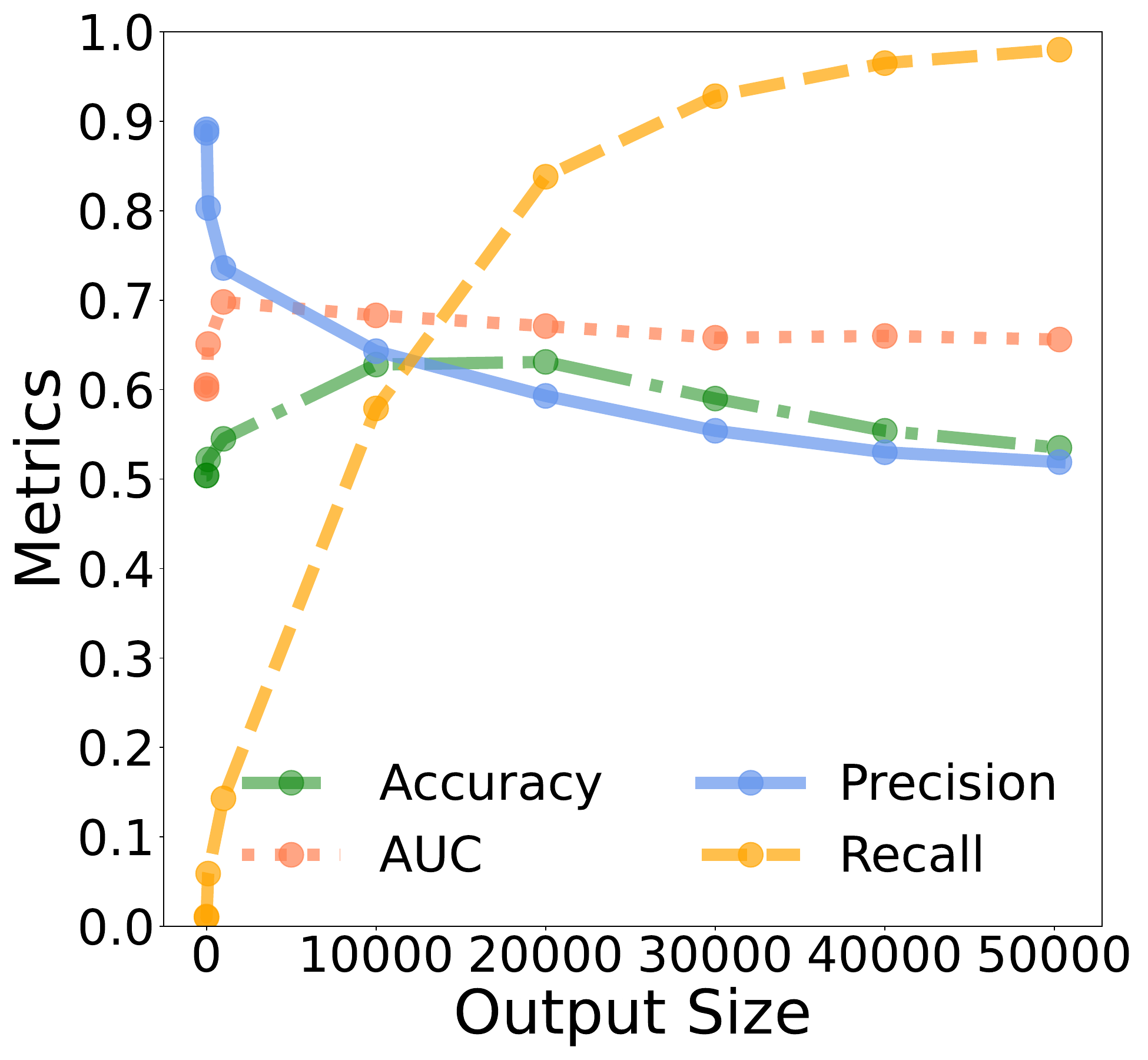}
	\caption{CodeGen}
	\label{fig_trend_b}
    \end{subfigure}
    \begin{subfigure}[b][][c]{.24\textwidth}
	\centering
        \includegraphics[width=\linewidth]{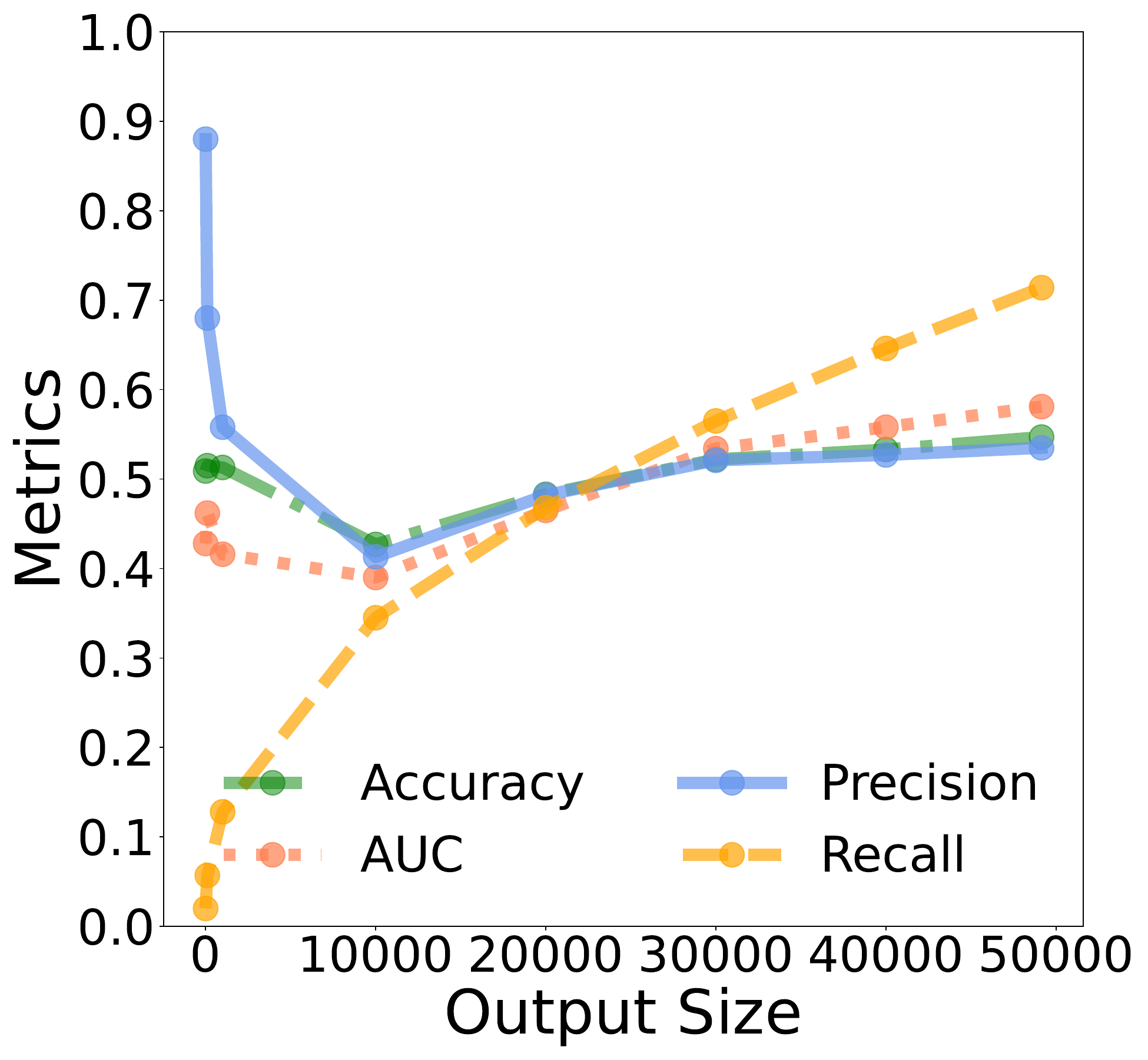}
	\caption{Starcoder}
	\label{fig_trend_b}
    \end{subfigure}
    \caption{The impact of changes in the output size of target models on membership inference.}
\label{fig_8}
\end{figure}
\begin{figure}[!t]
    \begin{subfigure}[b][][c]{.24\textwidth}
	\centering
        \includegraphics[width=\linewidth]{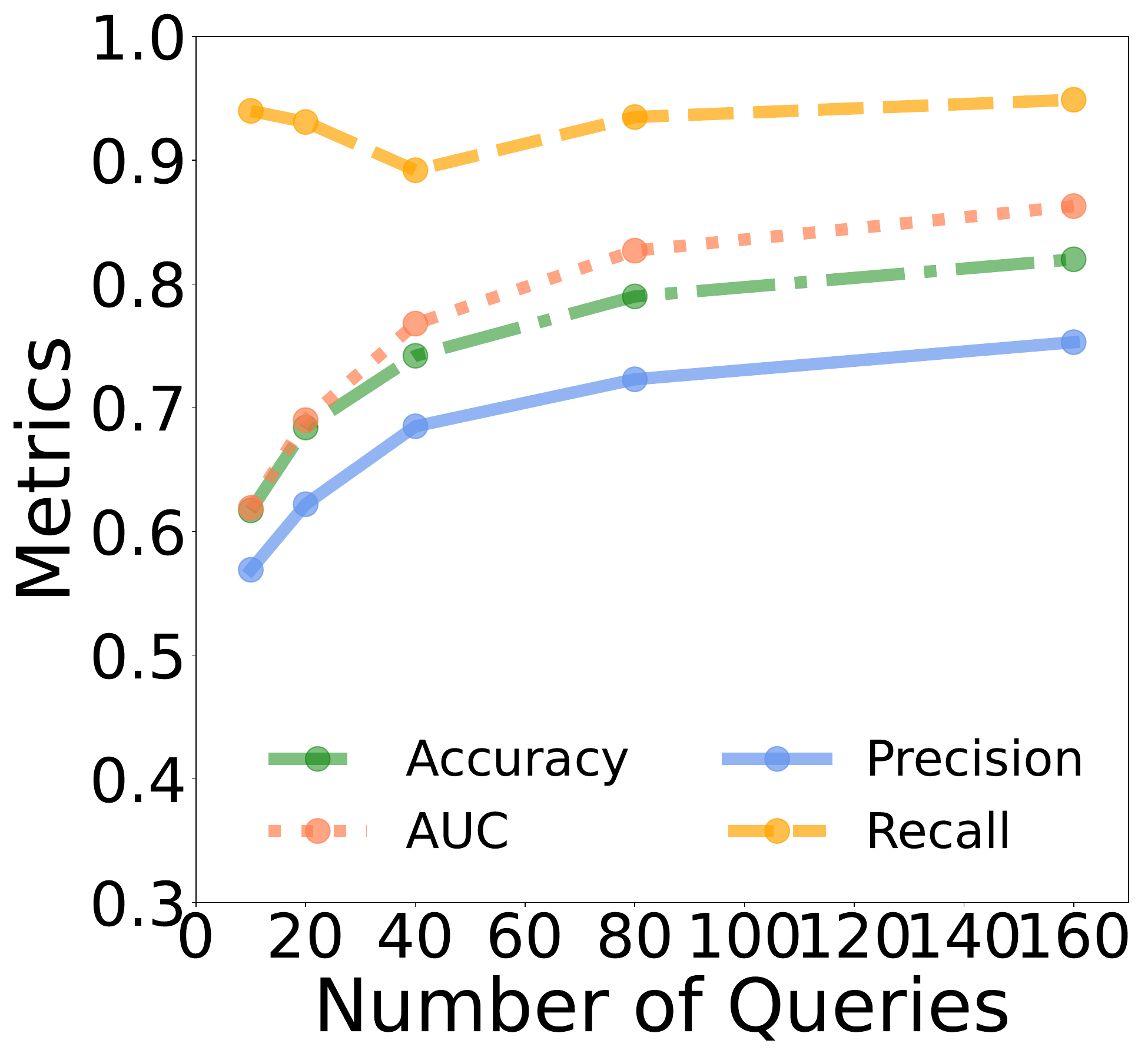}
	\caption{Random (LSTM)}
	\label{fig_trend_a}
    \end{subfigure}
    \begin{subfigure}[b][][c]{.24\textwidth}
	\centering
        \includegraphics[width=\linewidth]{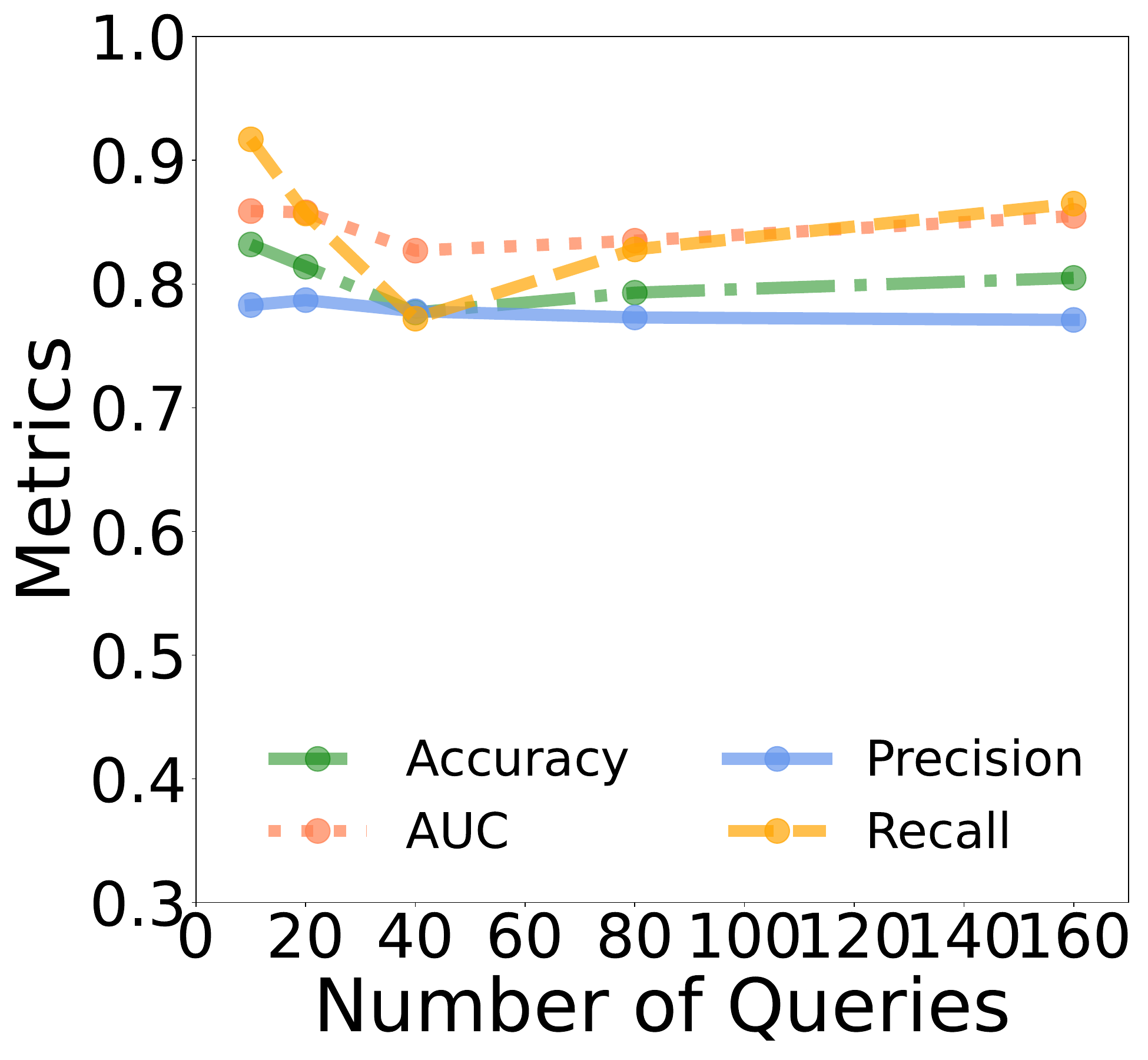}
	\caption{Frequency (LSTM)}
	\label{fig_trend_c}
    \end{subfigure}
    \begin{subfigure}[b][][c]{.24\textwidth}
	\centering
        \includegraphics[width=\linewidth]{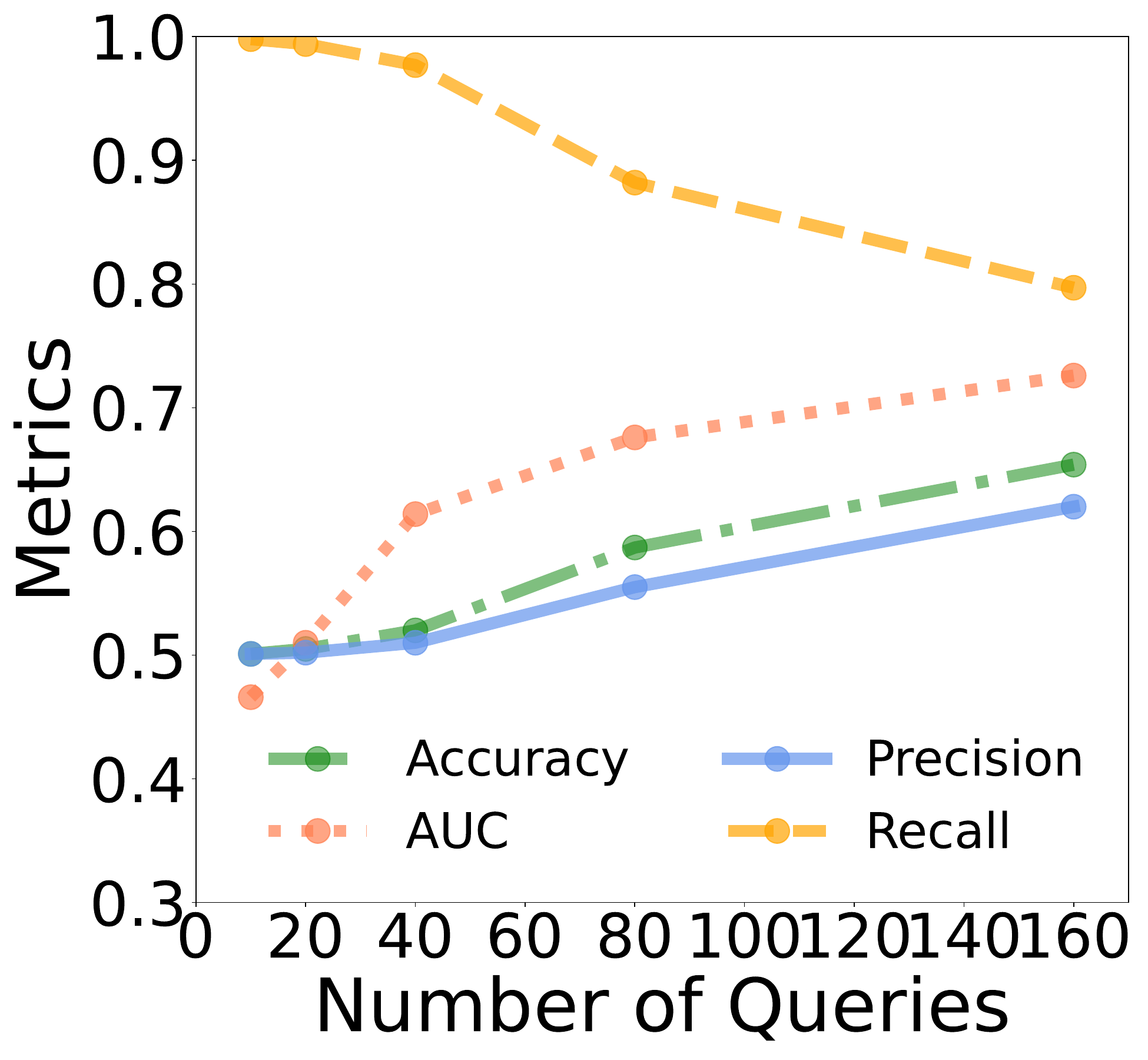}
	\caption{Random (CodeGPT)}
	\label{fig_trend_c}
    \end{subfigure}
    \begin{subfigure}[b][][c]{.24\textwidth}
	\centering
        \includegraphics[width=\linewidth]{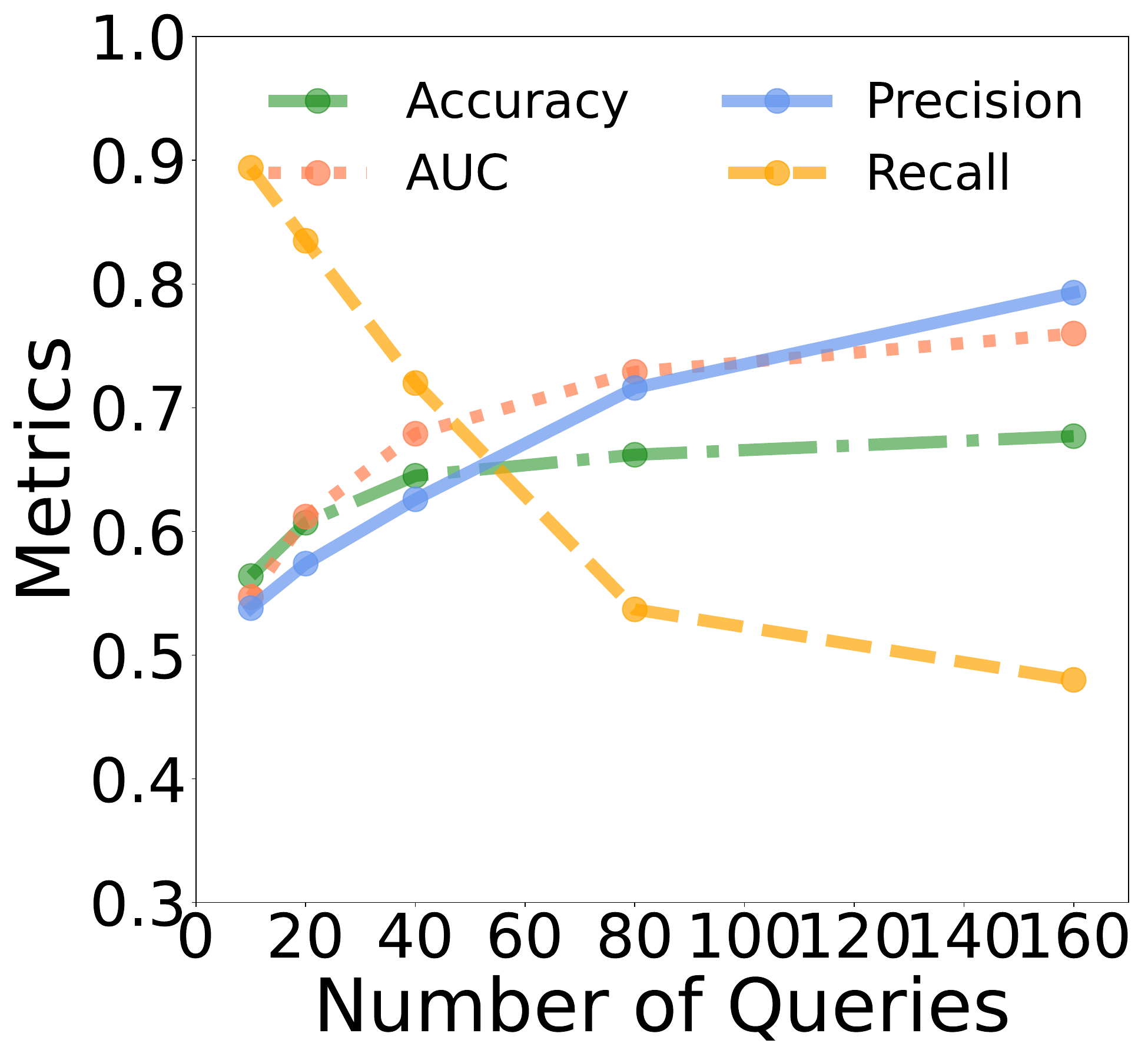}
	\caption{Frequency (CodeGPT)}
	\label{fig_trend_c}
    \end{subfigure}
    \begin{subfigure}[b][][c]{.24\textwidth}
	\centering
        \includegraphics[width=\linewidth]{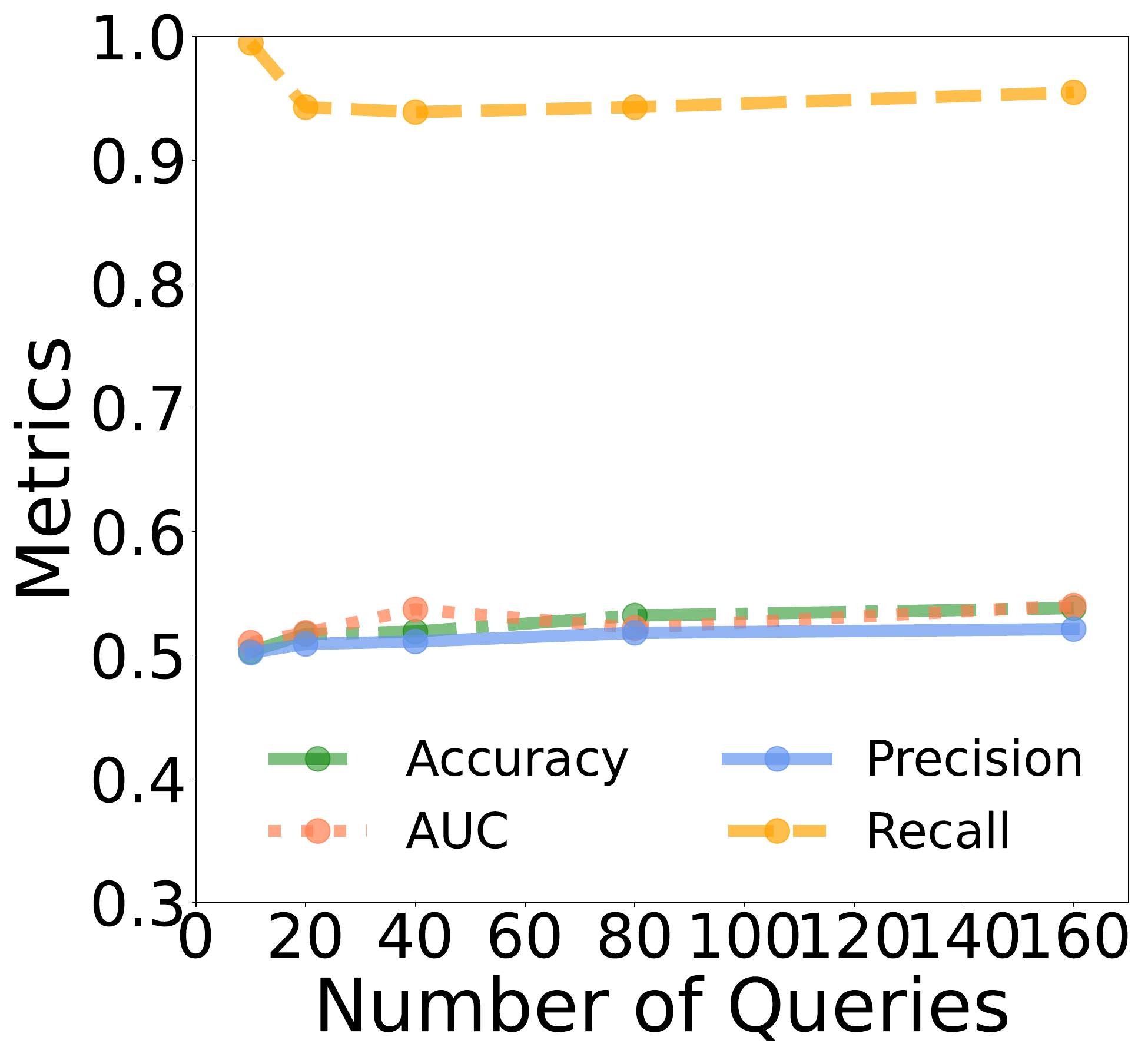}
	\caption{Random (CodeGen)}
	\label{fig_trend_c}
    \end{subfigure}
    \begin{subfigure}[b][][c]{.24\textwidth}
	\centering
        \includegraphics[width=\linewidth]{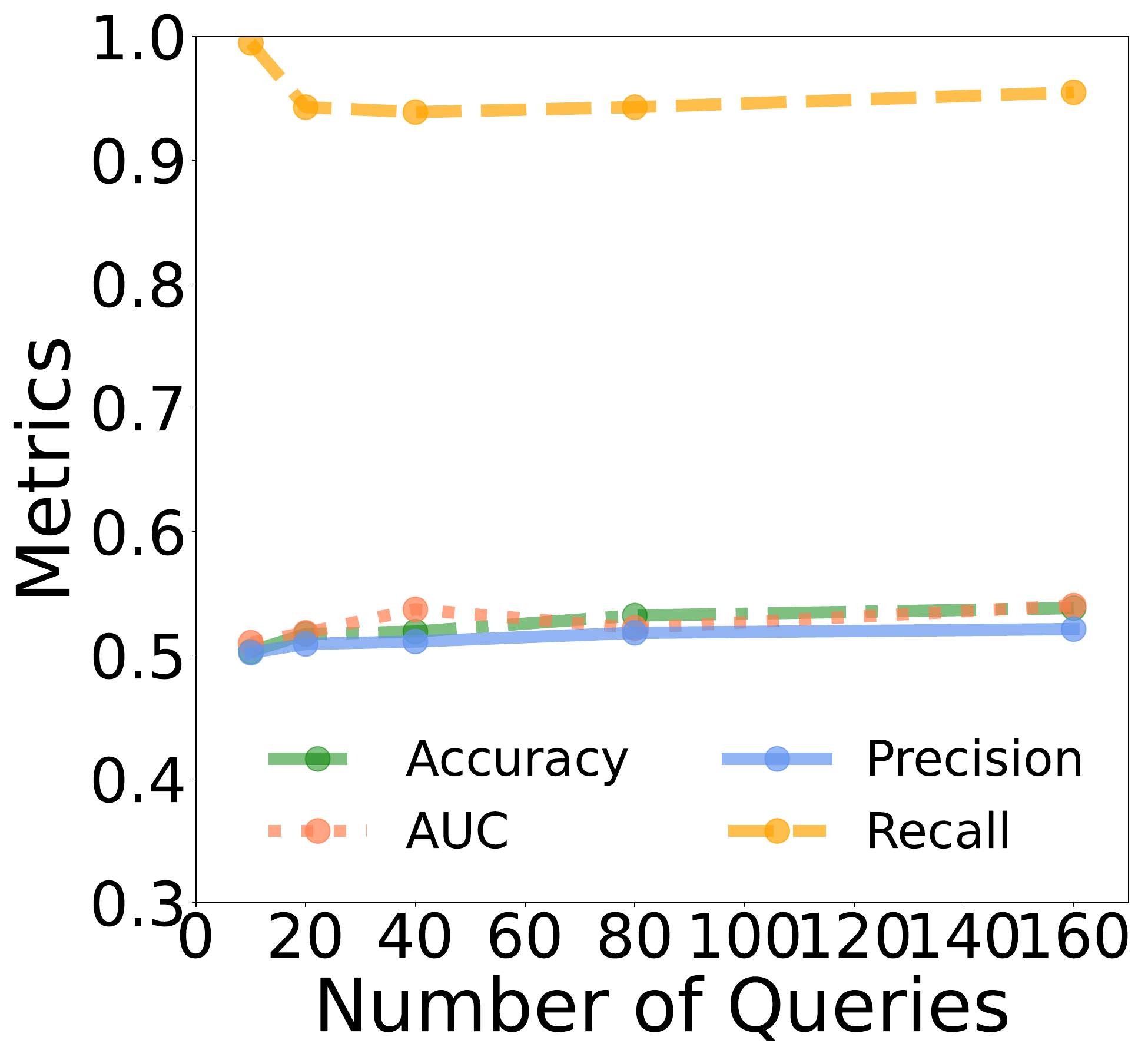}
	\caption{Frequency (CodeGen)}
	\label{fig_trend_c}
    \end{subfigure}
    \begin{subfigure}[b][][c]{.24\textwidth}
	\centering
        \includegraphics[width=\linewidth]{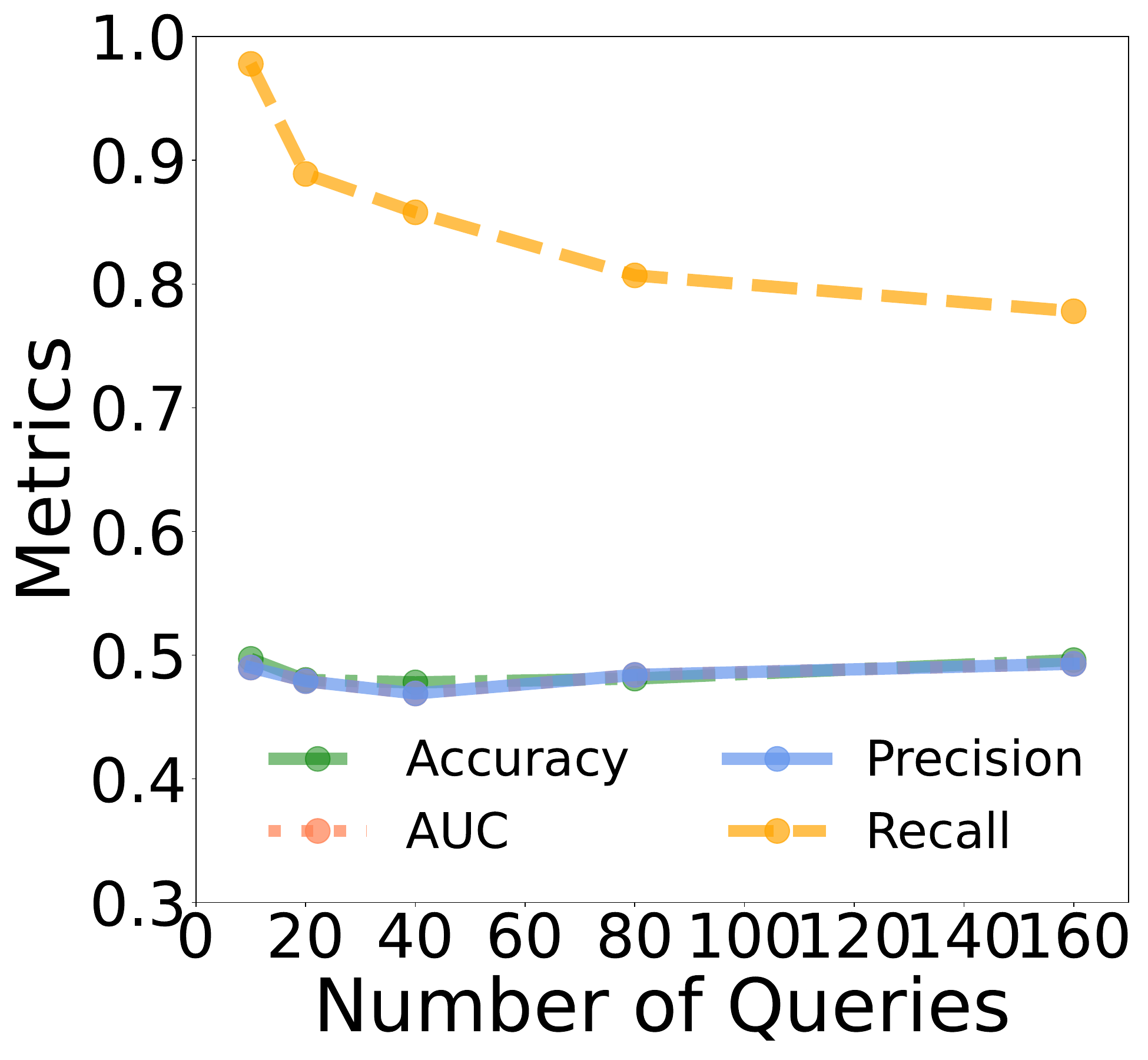}
	\caption{Random (StarCoder)}
	\label{fig_trend_c}
    \end{subfigure}
    \begin{subfigure}[b][][c]{.24\textwidth}
	\centering
        \includegraphics[width=\linewidth]{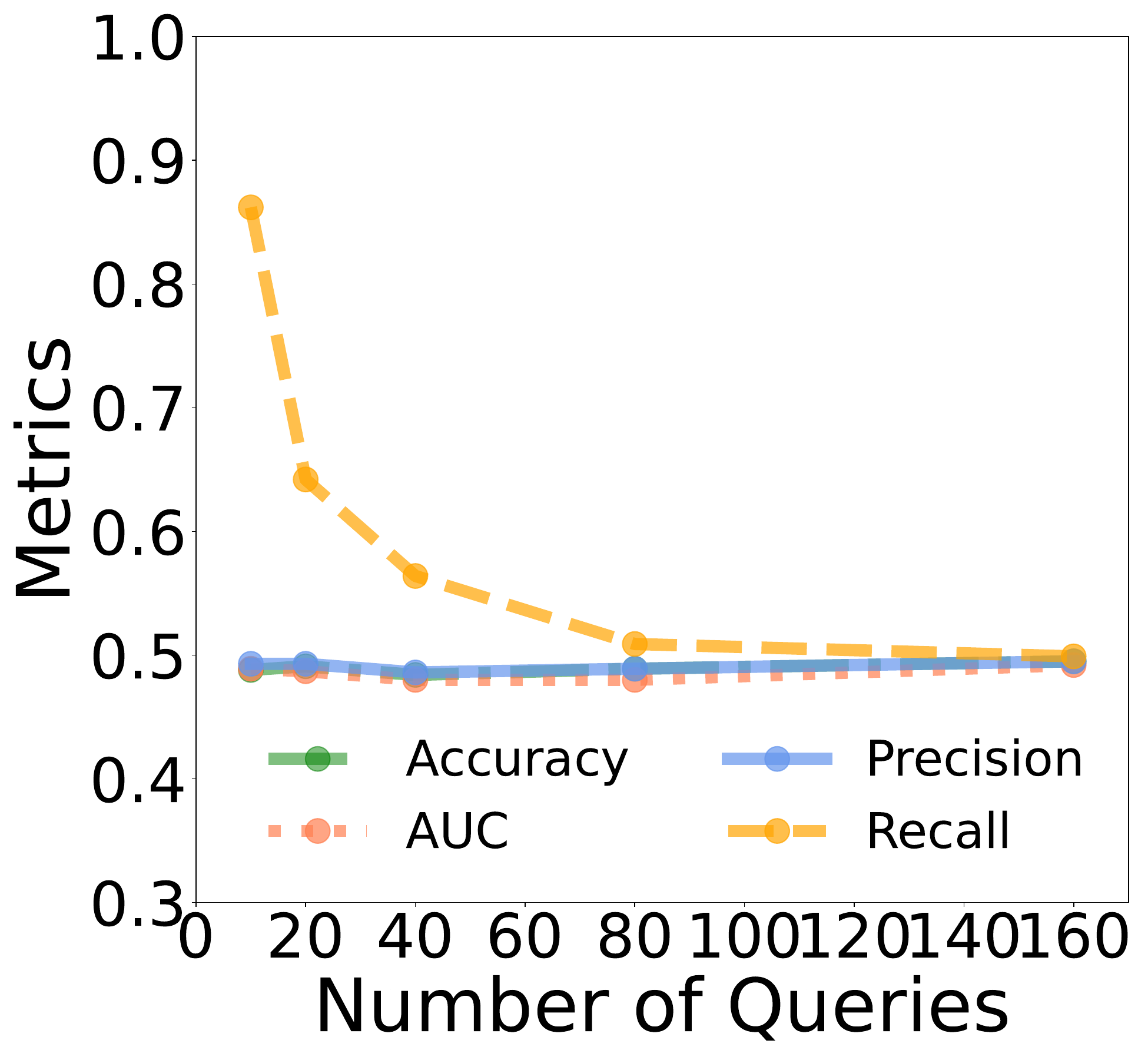}
	\caption{Frequency (StarCoder)}
	\label{fig_trend_c}
    \end{subfigure}
\caption{The performance of membership inference when varying the number of queries.}
\label{fig_9}
\end{figure}
\subsection{Impact of the Model's Output Size (RQ3)}\label{section_4.5}
Here, we investigate the  performance of \method across varying the output sizes of target models. 
More precisely, we systematically increase the output size of the target model, spanning from 1 to the entirety of the model's vocabulary size. Based on the reference, the vocabulary sizes for LSTM-based, CodeGPT, CodeGen, and StarCoder are 100,002, 50,234, 50,295, and 49,152, respectively.
Figure~\ref{fig_8} shows the results of membership inference targeting at four specific code completion models, while varying the model's output size. From this figure, we can see that as the output size of the models increases, a discernible enhancement in the performance of membership inference against LSTM-based and CodeGPT models becomes evident.
Moreover, the AUC for membership inference for target models except StarCoder shows acceptable performance when the output size of the model is set to 100 or 1000.
Overall, the effectiveness of our membership inference approach improves as the output size of the target model increases, benefiting from the utilization of more information. 
Nevertheless, when inferring a generated output from the target model becomes challenging, we observe that after a certain threshold in the target model's output size is reached, we can still achieve a remarkably successful membership inference. 
This phenomenon is can be attributed to the predominance of exceedingly high prediction ranks in the output.

\begin{tcolorbox}
	\textbf{Answer to RQ3.}  
 Generally, the more extensive the output size of the target code completion model is, the more pronounced the improvement in the performance of the membership inference model.
\end{tcolorbox}

\subsection{Impact of the Number of Queries (RQ4)}
\label{sec4.6}

Here, we investigate the influence of query quantity on the performance of \methodnospace.
To be more specific, we systematically vary the number of queries, under the range of 10, 20, 40, 80, and 160.
When we specify a count of 10 queries, it signifies that only a maximum of 10 queries are permissible for input into the target model for code completion. Our investigation revolves around two distinct strategies for constraining the number of queries:
1) Randomly sampling partial queries, and
2) Opting for queries characterized by the lowest frequency counts in their associated labels, as sequences containing relatively rare tokens tend to be more easily discernible.
Figure~\ref{fig_9} shows the performance of \method while varying the number of queries. 
Regarding the random sampling strategy, it becomes evident that membership inference performs poorly when fewer than 40 queries are made. However, as the number of queries increases to 160, we observe comparable performance with an Accuracy of 0.820 and 0.654 for LSTM-based and CodeGPT models, respectively.
Furthermore, with the frequency-based sampling strategy, even with only 20 queries for inference on the target model, we achieve an Accuracy of 0.814and 0.607 for LSTM-based and CodeGPT models, respectively.
These results demonstrate the flexibility and cost-effectiveness of our approach. Moreover, they show that code completion models exhibit distinct behavior between member and non-member data, particularly in the presence of rare words.

\begin{tcolorbox}
\textbf{Answer to RQ4.} Our proposed \method needs more queries when using the random sampling strategy, but with the frequency-based sampling strategy, comparable performance can be achieved with fewer queries. 
\end{tcolorbox}

\subsection{Case Study}
\begin{figure*}[!t]
\centering
\includegraphics[width=0.96\textwidth]{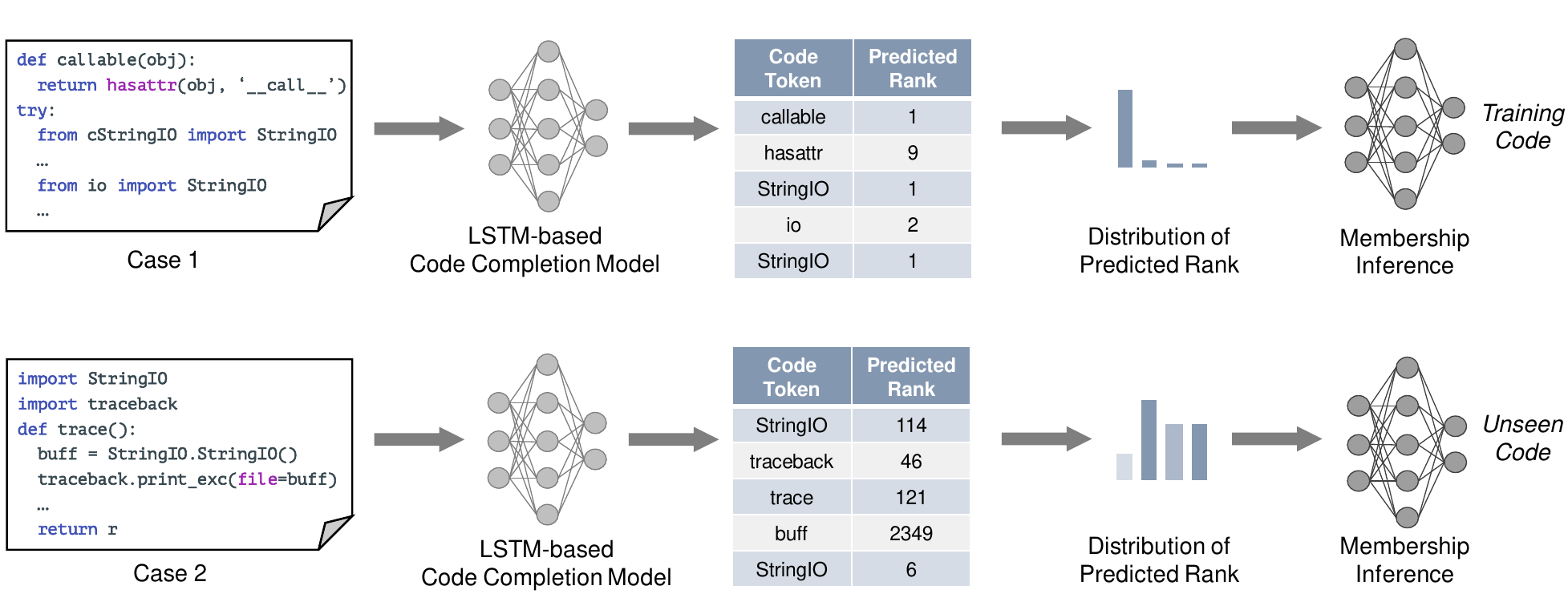}
\caption{Two real-world cases used to illustrate the effectiveness of \method.
}
\label{fig_8:case_studies}
\end{figure*}

To gain a deeper insight into our membership inference approach \methodnospace, we present two illustrative cases from our evaluation, as depicted in Figure~\ref{fig_8:case_studies}. In both of these cases, we configure the target code completion model as an LSTM-based model.
In the context of \textit{Case 1}, the code snippet showcased in the figure serves as the basis for training the corresponding code completion model. We employ this model to query partial code segments, generating its suggestions. To present the outcomes, we leverage predicted rankings, in accordance with the description in Section~\ref{sec_membership_classifier}. It's important to note that, due to space constraints, we can only provide a limited set of predicted tokens from the code completion model.
Examining this figure, we discern that the code completion model adeptly predicts the ground truth token. As an example, it confidently ranks \texttt{callable} as the top-1 suggestion.

Moving on to \textit{Case 2}, this case presents an example of code that was not encountered during the training phase of the code completion model. We observe that the code completion model tends to assign lower ranks (e.g., \#46 for the token \texttt{traceback}) in such scenarios.
Furthermore, an intriguing observation emerges from this case: the same token, \texttt{StringIO}, receives disparate predicted ranks when it appears in training code versus unseen code. Specifically, \texttt{StringIO} achieves a rank of 1 when situated within the training code context, but ranks 114 and 6 when encountered in the context of unseen code. Our membership inference classifier effectively captures this distinction, enabling it to predict membership status with success.

\section{Discussion}\label{sed_discussion}

In this section, we conduct an in-depth analysis to understand the performance of our membership inference approach on LSTM-based and CodeGPT code completion models, as well as its challenges when applied to LLMs of code, such as CodeGen and StarCoder.

\subsection{Why Does \method Work?}
While the membership inference approach has proven effective with LSTM-based and CodeGPT models, there is a pressing need to investigate the underlying factors contributing to its success. 
Generally, this effectiveness can be attributed to two key factors: (1) the prevalence of overfitting within the target code completion model, as described in Section~\ref{sec7.1.1} and (2) the model's propensity to memorize infrequently occurring tokens from the training dataset, a phenomenon also described as counterfactual memorization in prior research~\cite{zhang2023counterfactual}, as described in Section~\ref{sec7.1.2}. 

\begin{table}[t]
	\caption{The performance gap between training and testing of the target model. }
	\label{table__target_performance}
        \setlength{\tabcolsep}{6pt} 
\begin{tabular}{l|c|c}
\thickhline
Model & \multicolumn{1}{l|}{\textbf{Train Accuracy}} & \multicolumn{1}{l}{\textbf{Test Accuracy}} \\
\hline
LSTM-based & 0.707                           & 0.540                           \\
CodeGPT  & 0.767                            & 0.740  \\
\thickhline
\end{tabular}
\end{table}

\subsubsection{Overfitting of Target Code Completion Models}
\label{sec7.1.1}
The difference in the predictive performance of the target code completion model for member and non-member data is precisely the underlying principle of all membership inference method~\cite{shokri2017membership}, and the overfitting properties of the models are a visualization of this difference.
We assess the degree of overfitting in the target code completion models by comparing their training and testing accuracies.
Table~\ref{table__target_performance} shows the performance gap between training and testing, when targeting at the LSTM-based and CodeGPT models. 
Our experiments reveal that the LSTM-based model exhibits a substantial 16.7\% gap between training and testing accuracy, signifying poor generalization and a pronounced overfitting issue. 
Conversely, the CodeGPT model displays an impressively minimal train-test accuracy gap of under 3\%, underscoring its exceptional generalization and low susceptibility to overfitting which is one of the reasons why the membership inference approach on CodeGPT is not as good as that on LSTM-based model. 
This is particularly prominent in the LLMs of code such as CodeGen and StarCoder, which are trained on large-scale corpora with only 1 epoch, leading to their generalization very well. This leads to the fact that the effectiveness of membership inference attacks on LLMs of code needs to be improved.

\begin{figure}[!t]
    \begin{subfigure}[b][][c]{.24\textwidth}
	\centering
        \includegraphics[width=\linewidth]{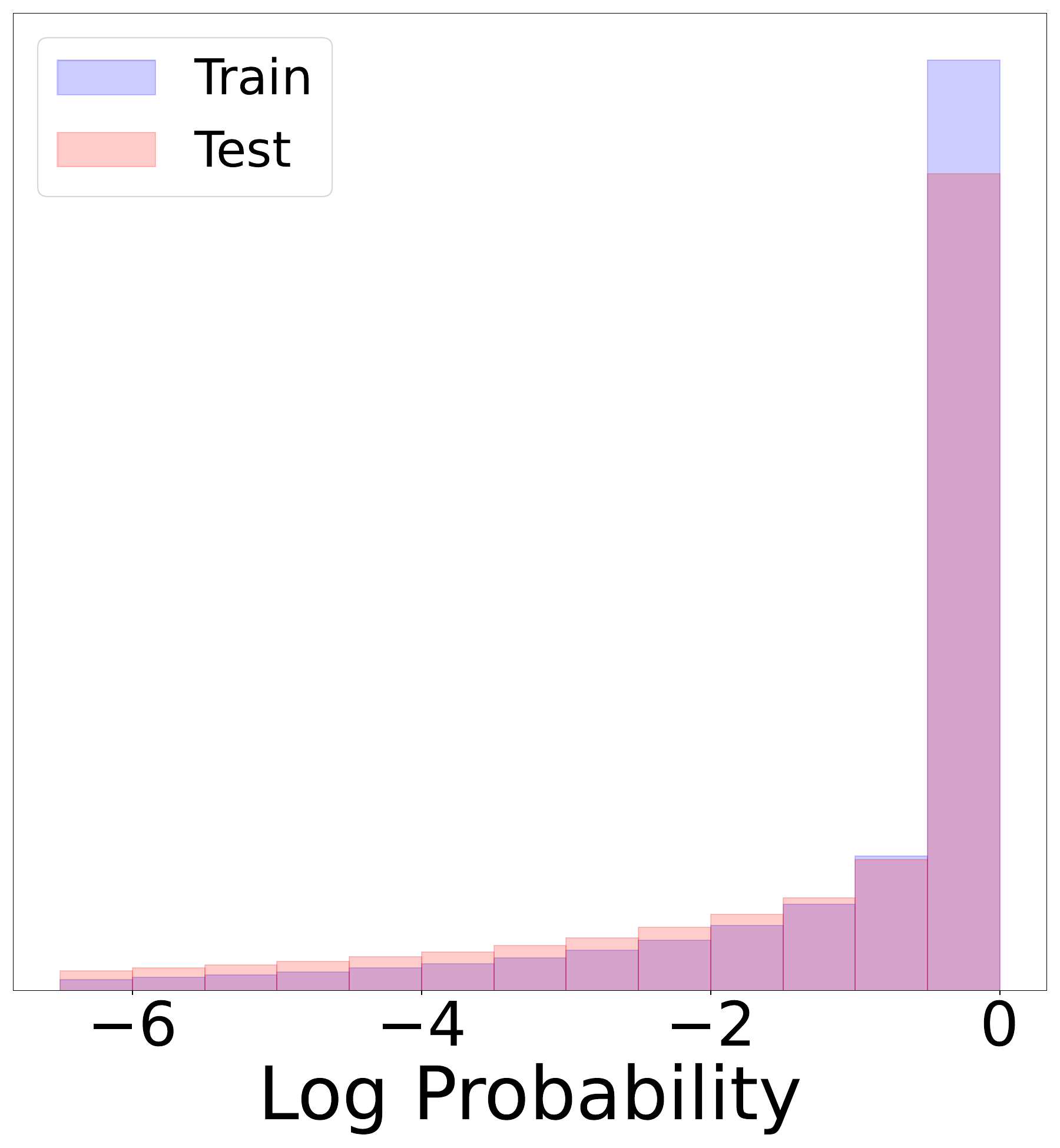}
	\caption{Top (LSTM-based)}
	\label{fig_logprob_a}
    \end{subfigure}
    \begin{subfigure}[b][][c]{.24\textwidth}
	\centering
        \includegraphics[width=\linewidth]{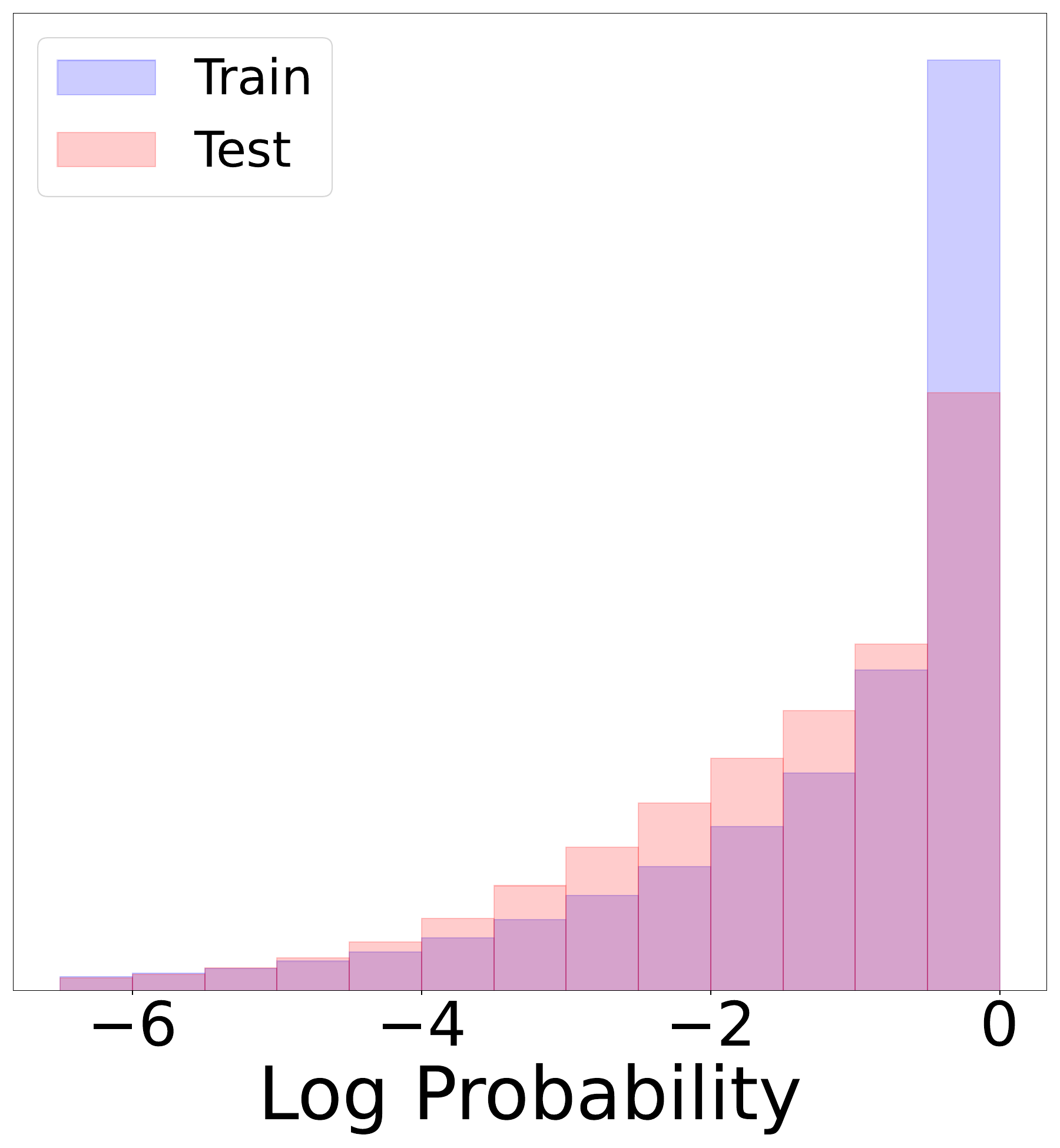}
	\caption{Tail (LSTM-based)}
	\label{fig_logprob_c}
    \end{subfigure}
    \begin{subfigure}[b][][c]{.24\textwidth}
	\centering
        \includegraphics[width=\linewidth]{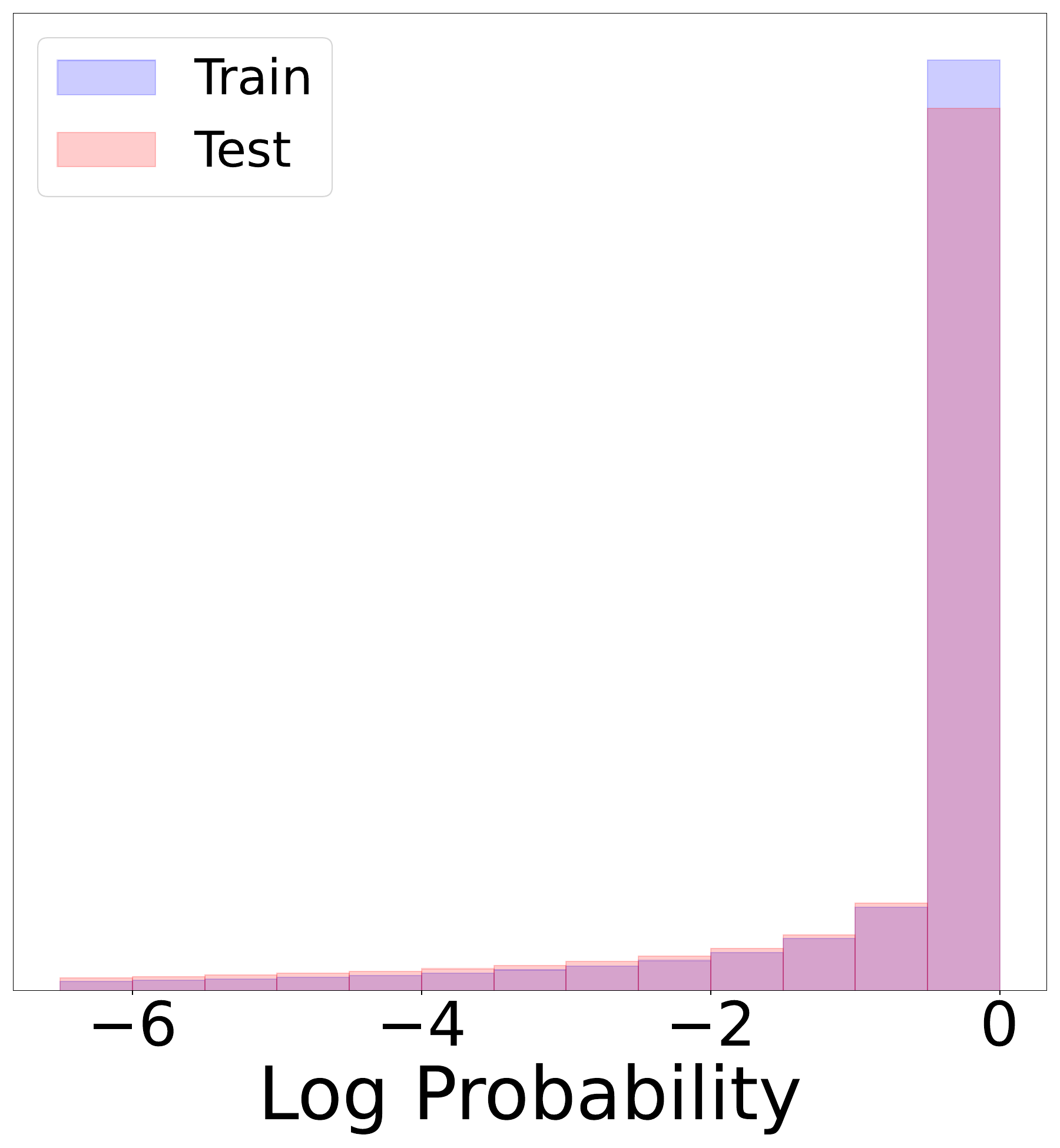}
	\caption{Top (CodeGPT)}
	\label{fig_logprob_c}
    \end{subfigure}
    \begin{subfigure}[b][][c]{.24\textwidth}
	\centering
        \includegraphics[width=\linewidth]{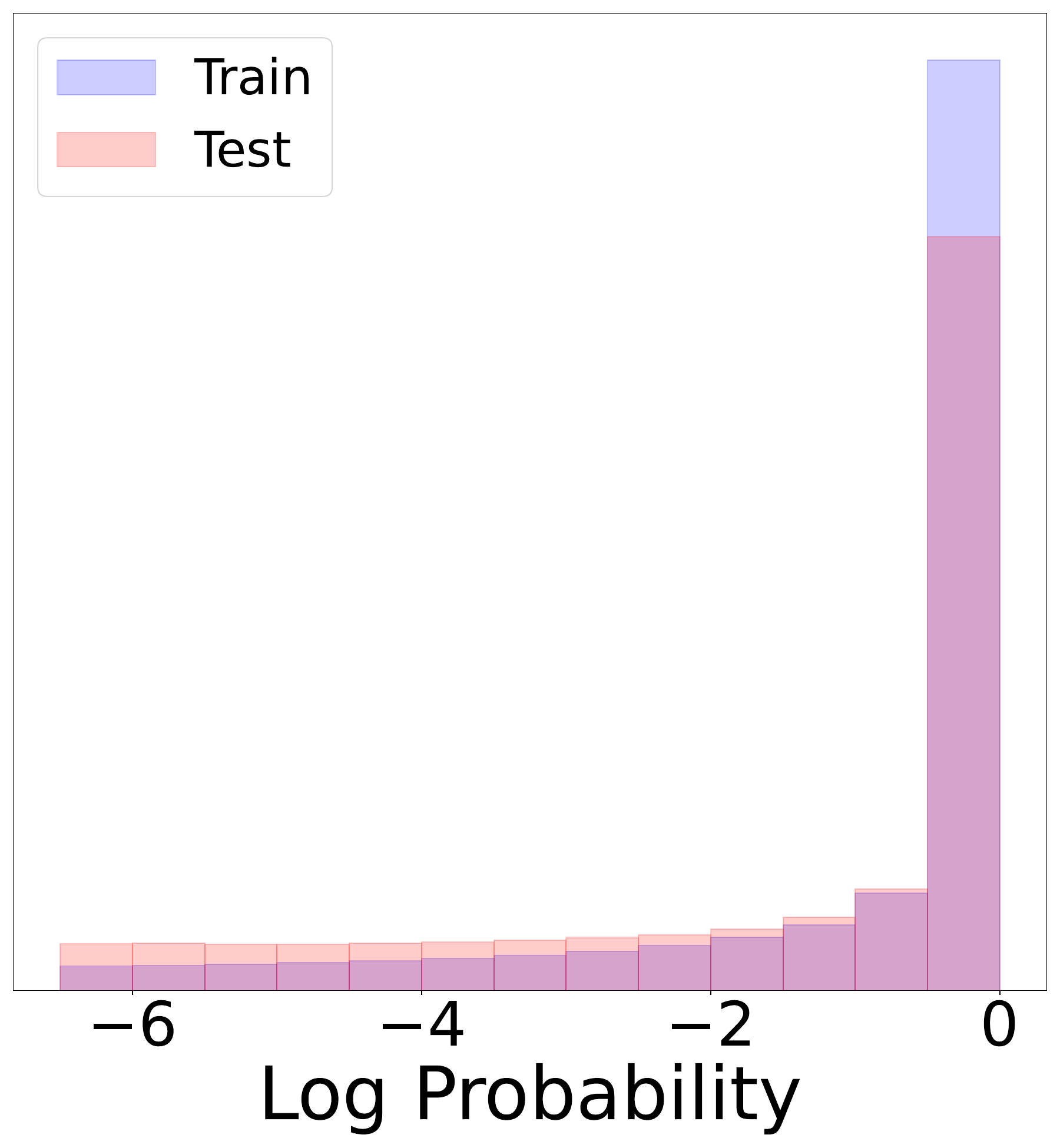}
	\caption{Tail (CodeGPT)}
	\label{fig_logprob_c}
    \end{subfigure}
    \begin{subfigure}[b][][c]{.24\textwidth}
	\centering
        \includegraphics[width=\linewidth]{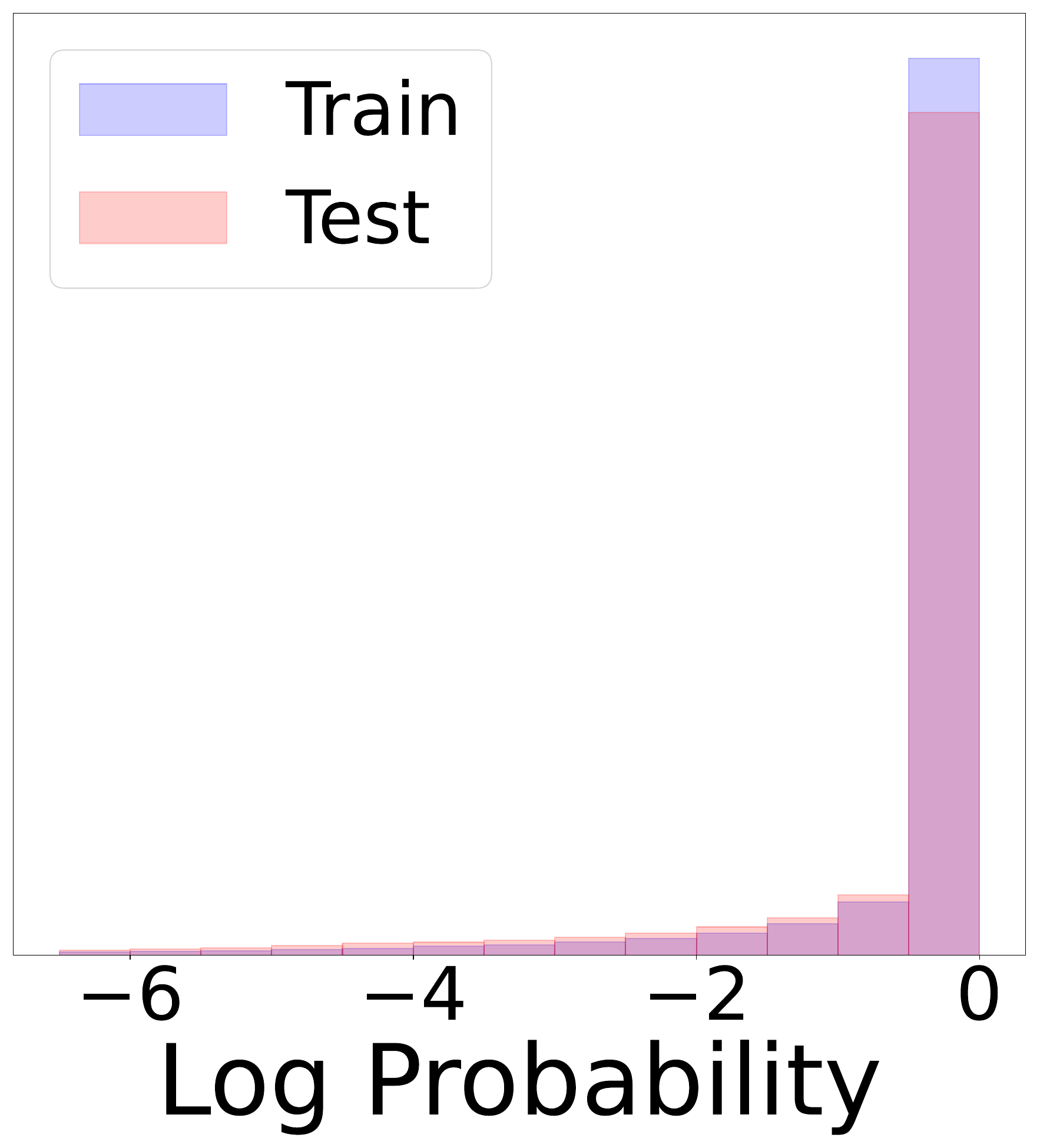}
	\caption{Top (CodeGen)}
	\label{fig_logprob_c}
    \end{subfigure}
    \begin{subfigure}[b][][c]{.24\textwidth}
	\centering
        \includegraphics[width=\linewidth]{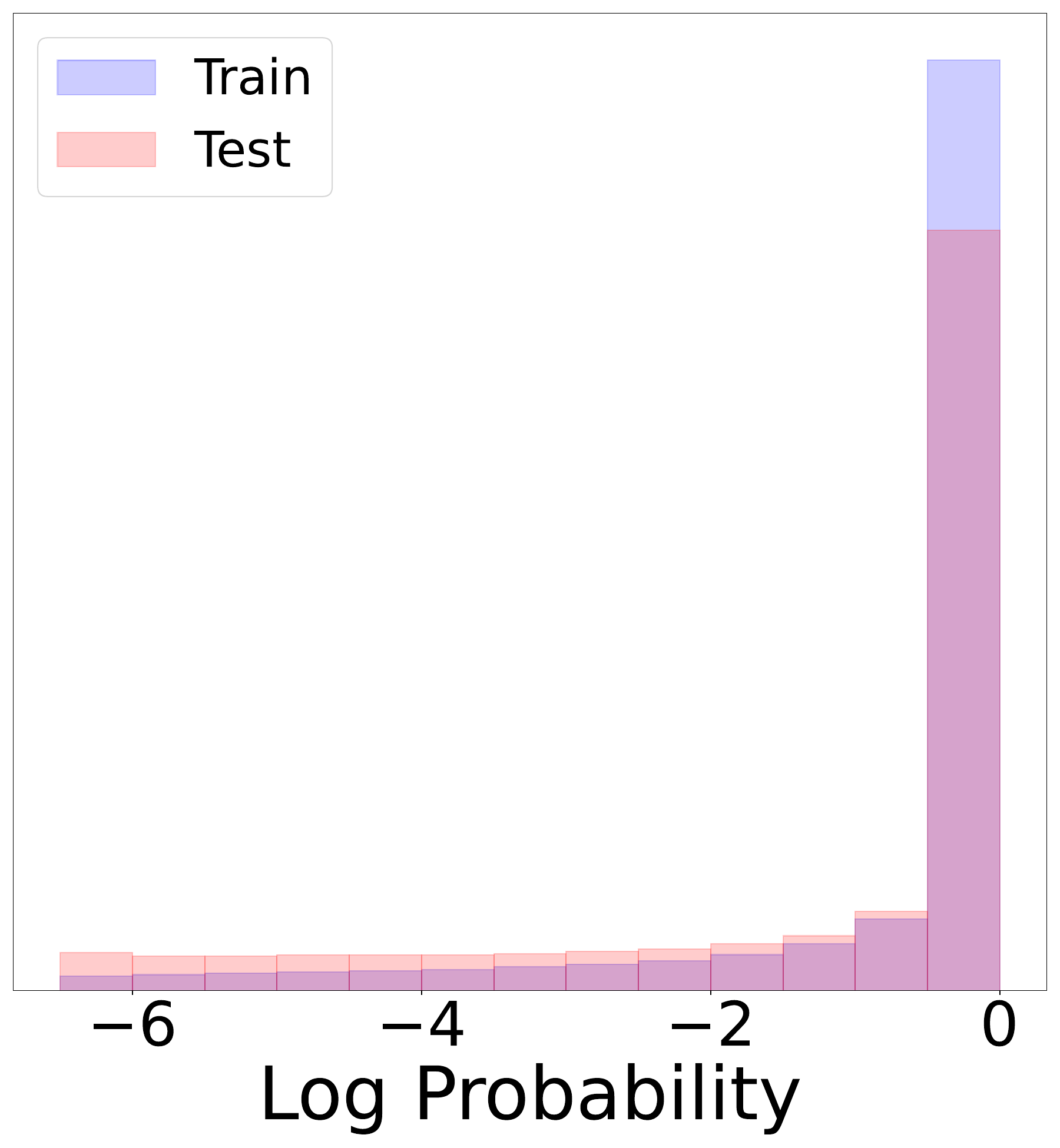}
	\caption{Tail (CodeGen)}
	\label{fig_logprob_c}
    \end{subfigure}
    \begin{subfigure}[b][][c]{.24\textwidth}
	\centering
        \includegraphics[width=\linewidth]{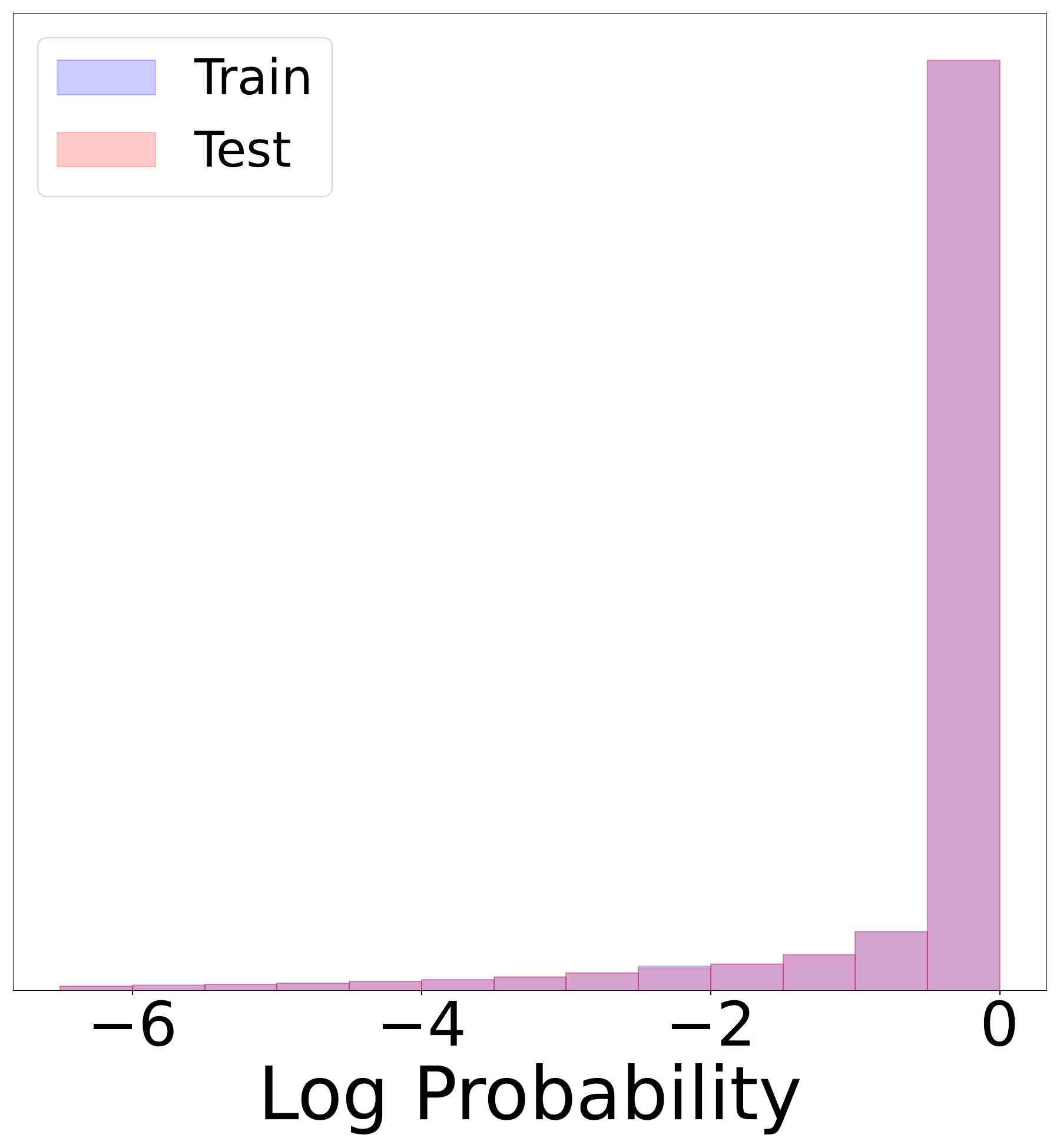}
	\caption{Top (StarCoder)}
	\label{fig_logprob_c}
    \end{subfigure}
    \begin{subfigure}[b][][c]{.24\textwidth}
	\centering
        \includegraphics[width=\linewidth]{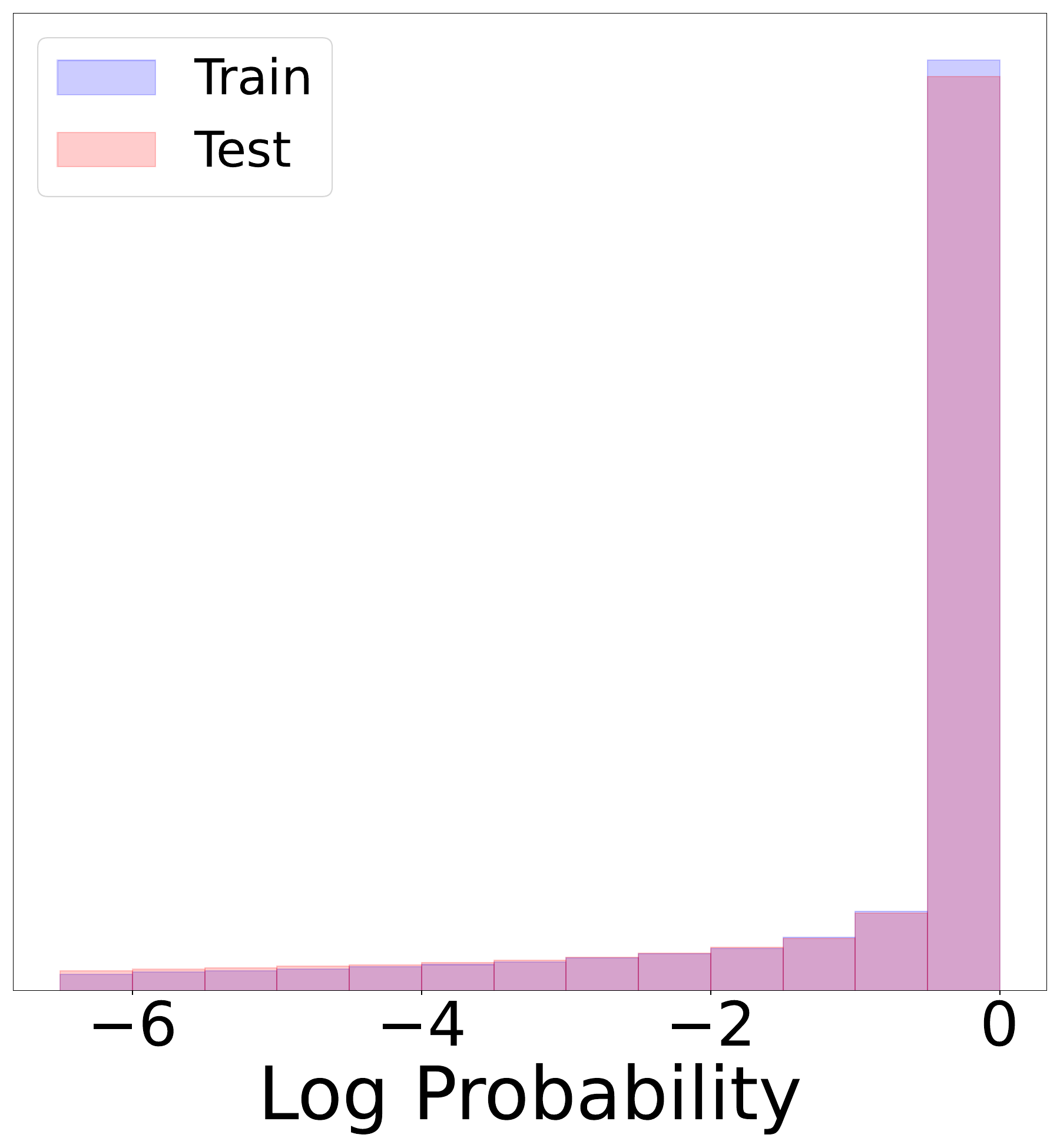}
	\caption{Tail (StarCoder)}
	\label{fig_logprob_c}
    \end{subfigure}
\caption{The histograms of log probabilities of tokens generated by the target neural code completion models (LSTM-based, CodeGPT, CodeGen and StarCoder). ``Top'' refers to histograms for the top 20\% most frequent tokens. ``Tail'' refers to the histograms for the rest.}
\label{fig_8:log_prob}
\end{figure}

\subsubsection{Model's Memory for Low-frequency Words} 
\label{sec7.1.2}
Another factor in the effectiveness of \method is the counterfactual memorization of the target models i.e. the memory for low frequency tokens. The model's specific memory for low-frequency training data will be exemplified in the following two perspectives.

\noindentparagraph{\textbf{\textup{Token Frequency \textit{v.s.} Predicted Probability.}}}
As previously discussed in Section~\ref{sec2.2}, our target code completion models employ a loss function based on negative log-likelihood. This loss function is designed to compel the model to commit to memory the sequences encountered in the training data. In Figure~\ref{fig_8:log_prob}, we present histograms depicting the logarithmic probability distribution of tokens generated by these neural code completion models.
Our statistical analysis led us to designate the code tokens appearing in the top 20\% frequency range in both the training and test datasets as ``Top-frequency'' tokens, constituting 89.05\%, 96.58\%, 95.86\%, and 95.16\% of the training datasets for LSTM-based, CodeGPT, CodeGen, and StarCoder, respectively. Conversely, all other code tokens are categorized as ``Tail-frequency''.
This figures underscores the fact that the code completion models consistently favor generating tokens from the set of code tokens with the highest frequency.

Moreover, we can deduce that when considering the disparity between the logarithmic probability distributions of the most frequently occurring tokens in both the training and test datasets, LSTM-based model exhibits a more pronounced dissimilarity compared to CodeGPT and CodeGen. Nevertheless, it is important to note that this dissimilarity, in the grand scheme of things, remains relatively modest.
Furthermore, when examining the LSTM-based model, a more conspicuous variation becomes apparent in the logarithmic probability distributions of the less common tokens within the training and test datasets. This variance implies that the LSTM-based model assigns higher probabilities to tokens within the training dataset, creating a robust signal that could potentially be exploited for membership inference.
However, in the case of the CodeGPT and CodeGen models, the disparity between the logarithmic probability distributions of less frequent tokens in the training and test datasets is comparatively slight. This discrepancy contributes to the less effective performance of membership inference for CodeGPT and CodeGen when contrasted with the LSTM-based model.
As for StarCoder, the difference in the logarithmic probability distributions between the training and test datasets is less pronounced. This observation might be a key factor contributing to StarCoder's suboptimal membership inference performance.

\begin{figure}[!t]
\centering
    \begin{subfigure}[b][][c]{.24\textwidth}
	\centering
        \includegraphics[width=\linewidth]{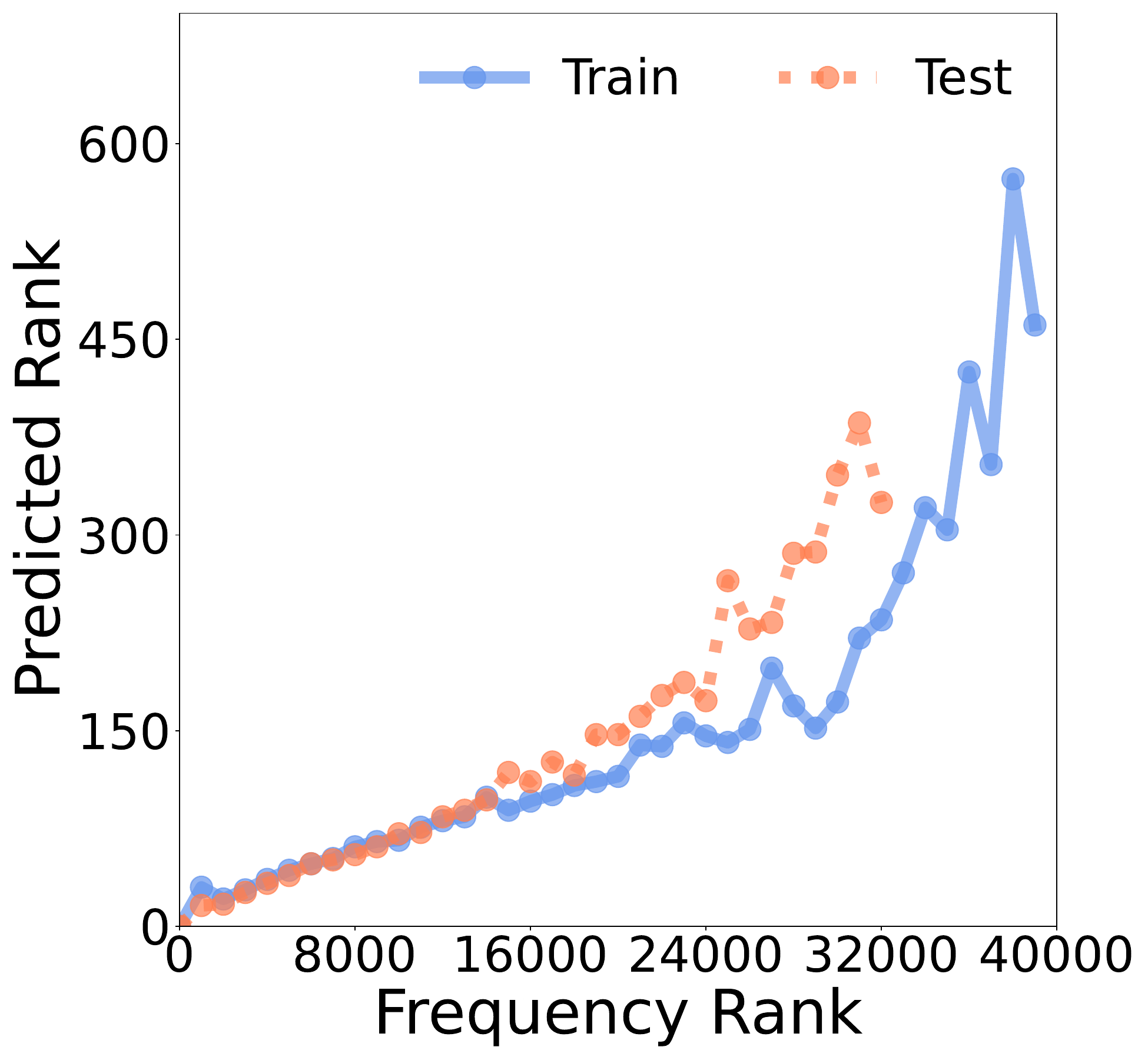}
	\caption{LSTM-based}
	\label{fig_trend_a}
    \end{subfigure}
    \begin{subfigure}[b][][c]{.24\textwidth}
	\centering
        \includegraphics[width=\linewidth]{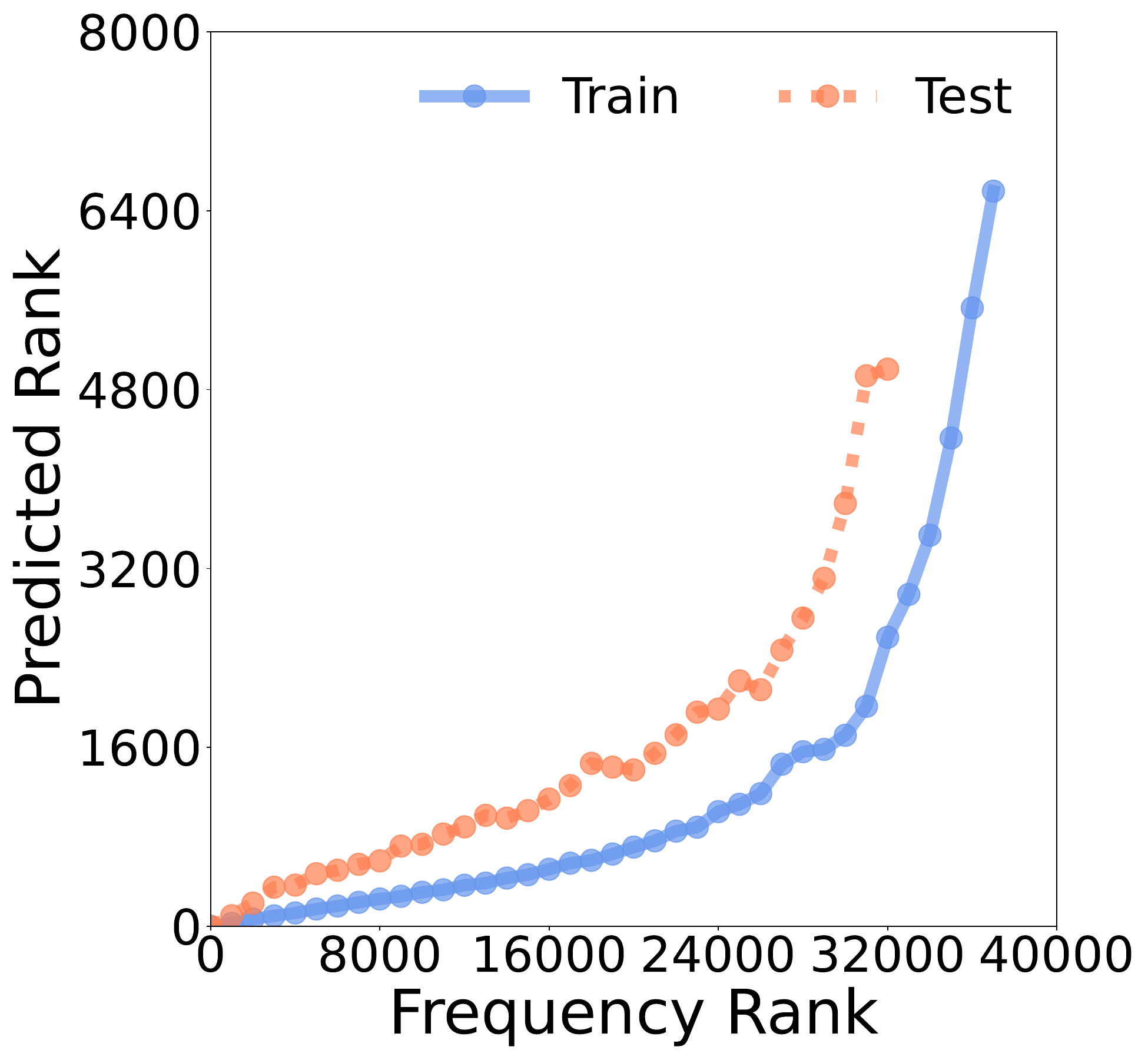}
	\caption{CodeGPT}
	\label{fig_trend_b}
    \end{subfigure}
    \begin{subfigure}[b][][c]{.24\textwidth}
	\centering
        \includegraphics[width=\linewidth]{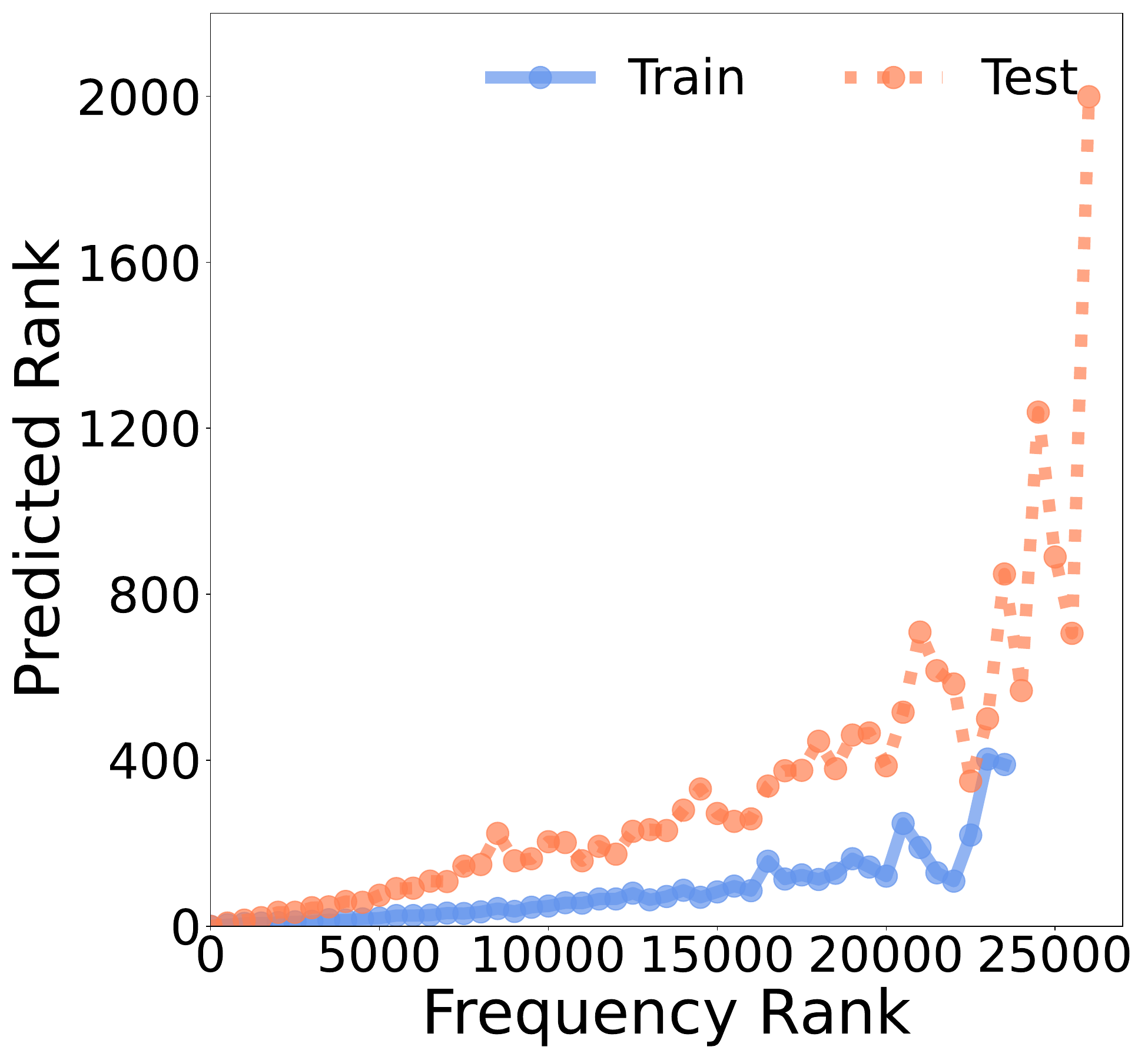}
	\caption{CodeGen}
	\label{fig_trend_b}
    \end{subfigure}
    \begin{subfigure}[b][][c]{.24\textwidth}
	\centering
        \includegraphics[width=\linewidth]{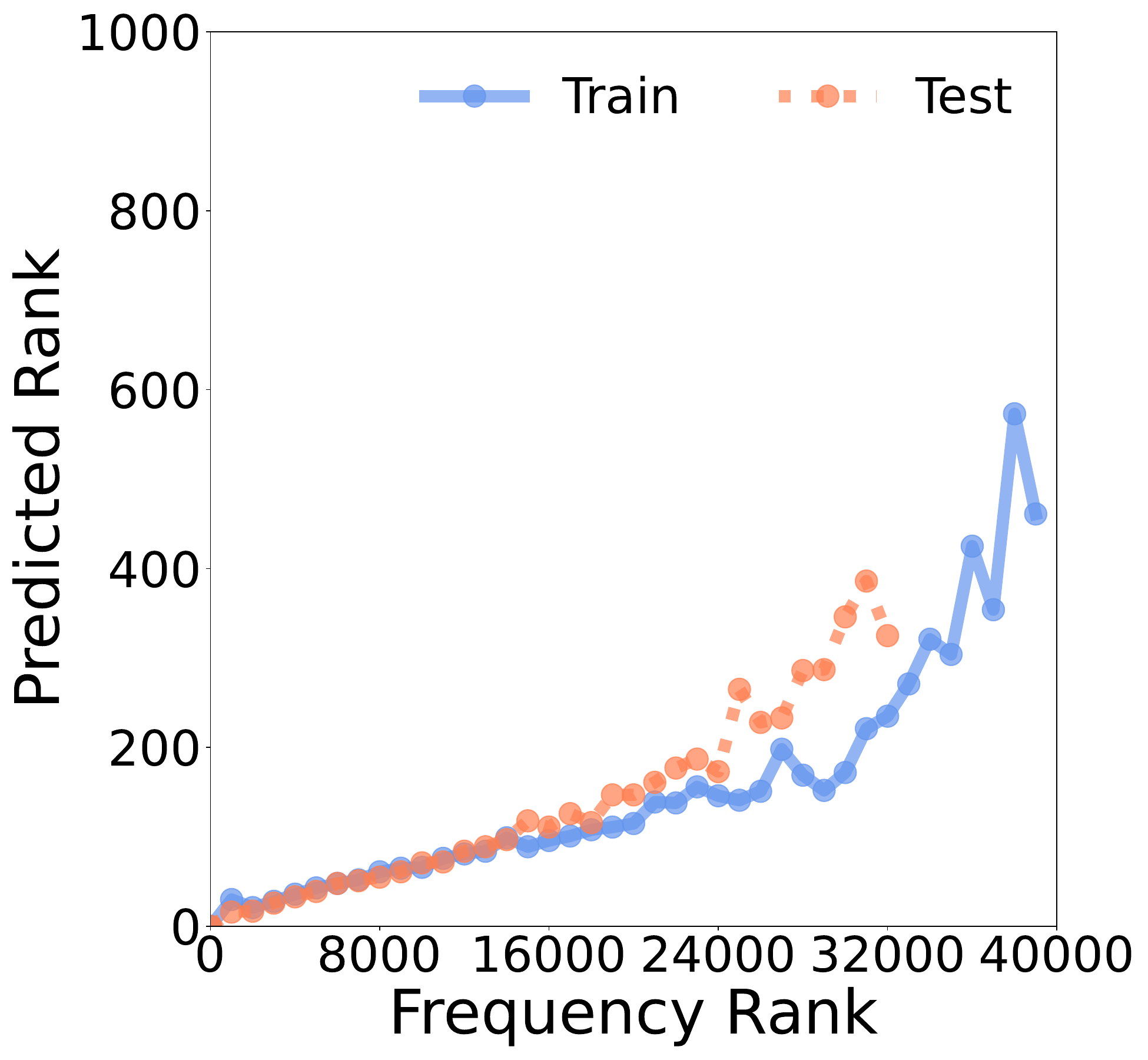}
	\caption{StarCoder}
	\label{fig_trend_b}
    \end{subfigure}
    \caption{
    The ranks of tokens in the frequency table of the training corpus and in the predictions generated by the models (where a lower rank signifies a higher probability for the token).
    }
\label{fig_12}
\end{figure}

\noindentparagraph{\textbf{\textup{Token Frequency \textit{v.s.} Predicted Rank.}}}
By committing the training sequence to memory, the model elevates the importance of tokens within that sequence when determining their relative ranking among candidate tokens in the output vocabulary. Figure~\ref{fig_12} illustrates the relationship between a token's rank in the training corpus's frequency table and its position in the model's predictions. A lower rank number signifies a higher token rank in the vocabulary, indicating greater frequency in the corpus or greater ease of prediction by the model.
In the case of these two models, less common tokens exhibit a notably wider gap between their model-predicted ranks when they appear in a training sequence versus a test sequence compared to more common tokens. This phenomenon partly accounts for why the performance of the membership inference approach improves as the model's output size increases (cf. Section~\ref{section_4.5}). With a larger output size, the model's output encompasses a greater number of low-frequency tokens, offering a more comprehensive reflection of the code completion model's memory concerning the training sequence.

\subsection{Further Discussions on Memorization Issues of LLMs}

Recent LLMs trained on extensive corpora exhibit exceptional generalization capabilities, effectively mitigating the risk of overfitting.
Consequently, this characteristic poses a unique challenge for our membership inference approach.
However, as highlighted in studies~\cite{carlini2022quantifying,carlini2021extracting,zhang2023counterfactual}, even LLMs that demonstrate proficient generalization may inadvertently memorize and subsequently disclose elements of their training dataset.
This phenomenon of memorization can be categorized into two distinct types: conventional memorization, which involves the retention of frequently occurring near-duplicate texts within a corpus, and counterfactual memorization, which pertains to the memorization of infrequent details from a particular training text~\cite{zhang2023counterfactual}.
\citet{carlini2022quantifying} explored the issue of memorization in LLMs, which falls into the former category of memorization types. 
Their study highlights a significant correlation: memorization becomes notably more pronounced in models with larger parameter sizes. Additionally, they identify another form of memorization characterized by the model's ability to retain information about low-frequency tokens, as previously described. These two aspects of memorization capacity underscore the feasibility of membership inference on LLMs.

\subsection{Threats to Validity}
There are several validity threats pertinent to our research.

\noindentparagraph{\textbf{\textup{On the ensurance of non-member dataset.}}}
The primary challenge to validity lies in our inability to ensure that the non-member datasets utilized for CodeGen and StarCoder do not contain any elements from the training data. Although extensive preprocessing measures have been implemented (outlined in Section~\ref{sec5.1}), the inherent duplication and code reusability could introduce bias into the results of membership inference.

\noindentparagraph{\textbf{\textup{Limited target code completion models.}}}
Another threat to validity pertains to the  limited target code completion models we have investigated.
In this paper, we validate our membership inference approach on two typical code completion models (i.e., LSTM-based model and CodeGPT) and two LLMs of code (i.e., CodeGen and StarCoder).
We believe that our membership inference approach can be easily extended to other code completion and code generation models, e.g., ChatGPT~\cite{chatgpt}, Code Llama~\cite{rozière2023code}.
We leave the validation on other code models to future work.

\noindentparagraph{\textbf{\textup{Limited dataset for evaluation.}}}
The third potential threat to validity arises from the fact that, in our experiments only the Python dataset (i.e., Py150, CodeSearchNet and Github-Python) is adopted for evaluation. 
In our future work, we will evaluate our method on more datasets of code in different programming languages.

\section{Related Work}
We review previous research from the perspectives of neural code completion, membership inference and dataset watermarking.

\subsection{Neural Code Completion}
Deep learning approaches have been widely used by researchers recently for code completion~\cite{raychev2014code,aye2020sequence,Zhangsurvey}. 
The first attempt to use neural language models in conjunction with program analysis to improve code completion was done by~\citet{raychev2014code}. Through program analysis, it first extracts the abstract histories of programs, and then it uses an RNN-based neural language model to learn the probabilities of histories.
Similar to this, a number of researches~\cite{liu2016neural,li2017code,svyatkovskiy2019pythia} use an RNN-based neural language model to infer the next code token over the incomplete AST after first traversing it in depth-first order.
\citet{kim2021code} proposed feeding the ASTs to Transformers in order to predict the missing partial code and better describe the code structure. 
A structural model for code completion was introduced by~\citet{alon2020structural}. It depicts code by sampling pathways from an incomplete AST.
Additionally, \citet{wang2021code} suggested utilizing Gated Graph Neural Networks~\cite{li2015gated} to represent the flattened sequence of an AST by parsing it into a graph.
IntelliCode Compose, a pre-trained language model of code based on GPT-2, was proposed by~\citet{svyatkovskiy2020intellicode} and offers rapid code completion across a variety of programming languages.
A multi-task learning framework that integrates the code completion and type inference tasks into a single overall framework was suggested by~\citet{liu2020self,liu2020multi}.
A retrieval-augmented code completion technique was presented by~\citet{lu2022reacc}. It gathers comparable code snippets from a code corpus and utilizes them as external context.
Recently, many pre-trained LLMs have showcased remarkable performance in code completion, including ChatGPT~\cite{chatgpt}, CodeGen~\cite{nijkamp2023codegen}, CodeT5+~\cite{wang2023codet5}, Code Llama~\cite{rozière2023code}, and StarCoder~\cite{li2023starcoder}.

\subsection{Membership Inference}
Membership inference attacks (MIAs) against machine learning models endeavor to discern whether a given data sample was utilized in the training of a target model or not.
Membership inference methods can be broadly divided into two categories, {i.e.,} shadow model training and metric-based techniques~\cite{hu2022membershipsurvey}.
Shadow model training~\cite{shokri2017membership} aims to train a binary attack classifier by creating multiple shadow models to mimic the behavior of the target model. 
Metric-based techniques~\cite{yeom2018privacy,song2021systematic,salem2019ml} proposed to infer the membership of a given data record by comparing the metric value ({e.g.,} prediction loss~\cite{yeom2018privacy}) calculated on the prediction vector of the given record to a preset threshold. 
Since both techniques are formulated from the attackers' standpoint, they assume the presence of abundant prior information, including knowledge of training data distribution and target model architectures. This assumption renders the direct application of these techniques unfeasible for data owners aiming to implement membership inference.
Recently, \citet{hu2022membership} leveraged backdoor techniques to allow the data owner to achieve membership inference effectively without such prior information. 
Their approach enables individuals to carry out membership inference solely by utilizing black-box query access to the target model.
One related work to ours is \cite{yang2023gotcha}, which also investigates the membership leakage risks of code models through membership inference and has been released as a preprint. However, our work diverges from this study in three key aspects. 
First, in experiments, we extend the scope beyond small code models, delving into the feasibility of implementing membership inference on LLMs such as CodeGen and StarCoder to safeguard code intellectual property. 
Second, instead of relying on prediction probabilities, we utilize the distribution of ranks of ground truth in the model output as the feature for the membership classifier to learn. Third, we conduct a comprehensive analysis to scrutinize the factors contributing to the effectiveness of our membership inference method.

\subsection{Dataset Watermarking}
Watermarking is a popular technique used for protecting the intellectual property (IP) of datasets~\cite{singh2013survey}. 
Many digital watermarking techniques~\cite{haddad2020joint, wang2021faketagger, guan2022deepmih} have been proposed to preclude unauthorized users from utilizing the protected data. 
\citet{mothi2019protection} introduced an innovative hybrid watermarking strategy that seamlessly merges visual cryptography with the Wavelet Packet Transform (WPT). This approach not only enables image segmentation but also facilitates the identification of the lowest energy band where the watermark is embedded.
Furthermore, aside from safeguarding the datasets, it is imperative to recognize that in the era of high-performing deep learning techniques, the trained models themselves should be regarded as intellectual property (IP) owned by their creators and accorded appropriate protection~\cite{li2021survey}.
Most recent works~\cite{nagai2018digital,qiao2023novel} aimed to plug the watermarks into the deep neural networks for easy authorization. 
However, unlike the above two directions, only a few prior works have started to protect illegal open-sourced data usage from deep neural network training.
Recently, \citet{sun2022coprotector} proposed an approach to protect the copyright of open-source code via data poisoning.
\citet{sun2023codemark} proposed CodeMark, to embed user-defined imperceptible watermarks into code datasets based on adaptive semantic-preserving transformations, to trace their usage in training neural code completion models.

\section{Conclusion}
This paper aims to explore the legal and ethical issues of current neural code completion models, which have not been systematically studied  previously.
We have put forward a membership inference approach. 
In particular, a series of shadow models are trained to emulate the function of the target code completion model. The acquired posteriors from these shadow models are subsequently employed to train a membership classifier. This classifier can then deduce the inclusion status of a given code sample by analyzing the output produced by the target code completion model.
Extensive experiments are carried out to measure the effectiveness of our proposed membership inference approach on two representative relatively small-scale neural code completion models (i.e., LSTM-based and CodeGPT) and two large-scale language models of code (i.e., CodeGen and StarCoder).
The experimental results demonstrate the capability of our membership inference approach to accurately ascertain whether a particular code sample is a part of the model's training dataset, with an Accuracy of 0.842 and 0.730 targeting LSTM-based model and CodeGPT, respectively. As for large language models of code with stronger generalization capabilities, membership inference methods still need to be further explored.

\noindentparagraph{\textbf{\textup{Data Availability.}}}
All experimental data and source code used in this paper are available at \texttt{\url{https://github.com/CGCL-codes/naturalcc/tree/main/examples/code-membership-inference}}~\cite{wan2022naturalcc}.

\balance
\bibliographystyle{ACM-Reference-Format}
\bibliography{ref}

\end{document}